\documentclass[acmtocl]{acmtrans2m}

\usepackage{latexsym}
\usepackage{amssymb}
\usepackage{prooftree}
\usepackage{diagrams}

\newtheorem{theorem}{Theorem}[section]

\newtheorem{lemma}[theorem]{Lemma}
\newdef{definition}[theorem]{Definition}
\newdef{remark}[theorem]{Remark}

\def\BB{\mathbb{B}}
\def\NN{\mathbb{N}}
\def\ZZ{\mathbb{Z}}
\def\RR{\mathbb{R}}
\def\PP{\mathbb{P}}
\def\FPS{\mathsf{R}}

\def\Func#1{{\mathsf{#1}}}

\def\Dom#1{\Func{dom}(#1)}

\def\Ran#1{\Func{ran}(#1)}
\def\Graph#1{\Func{Graph}(#1)}
\def\Rel#1#2{#1 \leftrightarrow #2}
\def\Loop#1#2{\Func{loop}(#1, #2)}
\def\Arr#1#2{#1 \rightarrow #2}
\def\DArrow{\stackrel{{\textstyle{\rightarrow}}}{{\rightarrow}}}
\def\LowerStackRel#1#2{\mathrel{\mbox{\raisebox{-0.5ex}{$\stackrel{#1}{#2}$}}}}

\def\|{\mathrel{|}}
\def\CRel#1{\Func{Rel}(#1)}
\def\LSp#1{{^{#1}}}
\def\Norm#1{{||}#1{||}}

\newcommand{\fdefin}{:\equiv}

\newcommand{\pdefin}{:\equiv}

\newcommand{\eqleft}[1]{\begin{itemize} \item[] $#1$ \end{itemize}}

\newcommand{\eProof}{$\quad \Box$ \\}

\newcommand{\nops}{{\sf id}}
\newcommand{\assign}{:=}
\newcommand{\while}{{\sf while}}
\newcommand{\twist}{{\sf twist}}
\newcommand{\cond}[1]{\nabla_{\!#1}}
\newcommand{\inj}{{\sf inj}}
\newcommand{\linj}{\inj_0}
\newcommand{\rinj}{\inj_1}

\newcommand{\sCopy}{{\sf c}}
\newcommand{\sTl}{{\sf tl}}
\newcommand{\sR}{{\sf R}}

\newcommand{\Lang}{{\mathcal L}}
\newcommand{\fLang}{\Lang_{\mathbb{FC}}}
\newcommand{\pLang}{\Lang_{\mathbb{PP}}}
\newcommand{\bLang}{{\mathcal L}^b}

\newcommand{\catS}{{\cal S}}
\newcommand{\fcatS}{{\cal S}_{\mathbb{FC}}}
\newcommand{\tfcatS}{{\cal S}^T_{\mathbb{FC}}}
\newcommand{\pcatS}{{\cal S}_{\mathbb{PP}}}
\newcommand{\scatS}{{\cal S}_{\mathbb{SC}}}

\newcommand{\catSo}{{\cal S}_o}
\newcommand{\catSm}{{\cal S}_m}

\newcommand{\Hoa}{{\cal H}}
\newcommand{\fcH}{H_{\mathbb{FC}}}
\newcommand{\tfcH}{H^T_{\mathbb{FC}}}
\newcommand{\ppH}{H_{\mathbb{PP}}}
\newcommand{\scH}{H_{\mathbb{SC}}}

\newcommand{\Seq}[2]{#1; #2}
\newcommand{\Fb}{{\sf Fb}}
\newcommand{\Par}{{\sf Par}}

\newcommand{\Tr}{{\sf Tr}}

\newcommand{\pair}[2]{\langle #1, #2\rangle}

\newcommand{\iord}{\sqsubseteq}

\newcommand{\id}{{\sf id}}

\newcommand{\ifthen}[3]{(#1)?(#2, #3)}
\newcommand{\sm}[1]{[\![#1]\!]}
\newcommand{\rel}[3]{#1 \; \underline{#2} \; #3}

\newcommand{\hTriple}[3]{\{#1\} \; #2 \; \{#3\}}

\newcommand{\sPost}{{\sf SPC}}
\newcommand{\wPre}{{\sf WPC}}
\newcommand{\RRT}{{\sf RRT}}

\newcommand{\Store}{{\sf Store}}
\newcommand{\Heap}{{\sf Heap}}
\newcommand{\State}{{\sf State}}
\newcommand{\Var}{{\sf Var}}
\newcommand{\abort}{{\sf abort}}
\newcommand{\mwand}{-\!\!*\,}
\newcommand{\Hsystem}[3]{\textup{\textbf{HL}}(#1, \sm{\cdot}, #3)}
\newcommand{\EFL}{\textup{\textbf{EFL}}}
\def\Func#1{{\mathsf{#1}}}

\def\Dom#1{\Func{dom}(#1)}

\def\Ran#1{\Func{ran}(#1)}

\title{A General Framework for Sound and Complete Floyd-Hoare Logics}

\author{ROB ARTHAN, URSULA MARTIN, ERIK A. MATHIESEN and PAULO OLIVA \\
Queen Mary, University of London}

\begin{abstract} This paper presents an abstraction of Hoare logic to traced symmetric monoidal categories, a very general framework for the theory of systems. Our abstraction is based on a traced monoidal functor from an arbitrary traced monoidal category into the category of pre-orders and monotone relations. We give several examples of how our theory generalises usual Hoare logics (partial correctness of while programs, partial correctness of pointer programs), and provide some case studies on how it can be used to develop new Hoare logics (run-time analysis of while programs and stream circuits).
\end{abstract} 




\begin{document}

\begin{bottomstuff} This research was supported by EPSRC grants GR/M98340, GR/L48256 and EP/F02309X/1 and by an EPSRC studentship to the third author. Oliva also gratefully acknowledges support of the Royal Society under grant 516002.K501/RH/kk.  \newline Authors' address: Department of Computer Science, Queen Mary, University of London, Mile End Road, London E1 4NS, UK.
\end{bottomstuff}

\maketitle

\section{Introduction} 


Under the general label of \emph{Hoare logic}, the early work of  Floyd \citeyear{Floyd:1967} and Hoare \citeyear{Hoare:1969} on axiom systems for flowcharts and while programs  has been applied to various other domains, such as recursive procedures \cite{Apt:1981}, pointer programs \cite{Reynolds:2002}, and  higher-order languages \cite{Berger:2005}.  Our goal in this paper is to identify a minimal structure supporting soundness and (relative) completeness results, in the manner of  Cook's presentation for \emph{while programs} \cite{Cook:1978}. This is achieved via an abstraction of Hoare logic to the theory of traced symmetric monoidal categories \cite{Joyal:1996}, a very general framework for the theory of systems.

Traced symmetric monoidal categories  precisely capture the intrinsic structure of both flowcharts and dynamical systems, namely: sequential and parallel `composability', and feedback (unbounded iteration). In fact, traced symmetric monoidal categories are closely related to Bainbridge's work \citeyear{Bainbridge:1976} on the duality between flowcharts and networks. The scope of traced symmetric monoidal categories, however, is much broader, being actively used, for instance, for modelling computation (e.g. \cite{Simpson:2000}) and in connection with Girard's geometry of interaction (e.g. \cite{Haghverdi:2005}).

The main feature of Hoare logic is the use of assertions $P, Q$ specifying the input-output behaviour of a program. If $A$ is a program, then the triple $\hTriple{P}{A}{Q}$ states that on inputs satisfying $P$ the program $A$, if it terminates, will produce a result which satisfies property $Q$. An inherent ordering among assertions, given by the consequence relation, is also available, allowing for the properties of input/output to be refined or relaxed. In particular, if $\hTriple{P}{A}{Q}$ holds, and $Q$ logically implies $Q'$, then we must also have that $\hTriple{P}{A}{Q'}$ holds. Abstractly, assertions can be viewed as objects of a \emph{pre-ordered set}, with the logical implication as an instance of an ordering relation. The derived Hoare triple relation $\hTriple{\cdot}{A}{\cdot}$ can then be viewed as \emph{monotone} binary relation between the points of the pre-order.

More precisely, let $\Hoa$ be the category of pre-ordered sets and monotone relations (see Section \ref{abs-hoare} for formal definition). We first identify a particular class of functors -- which we call \emph{verification functors} -- between traced symmetric monoidal categories and subcategories of $\Hoa$. We then give an abstract definition of Hoare triples, parametrised by a verification functor, and prove a single soundness and completeness (in the sense of Cook) theorem for such triples. In the particular case of the traced symmetric monoidal category of while programs (respectively, pointer programs) this embedding gives us back Hoare's original logic \citeyear{Hoare:1969} (respectively, O'Hearn and Reynolds logic \citeyear{Reynolds:2002}). In order to illustrate the generality of our framework, we also derive new sound and complete Hoare-logic-like rules for the verification of running-time (and hence termination) of programs, and the verification of linear dynamical systems (modelled via stream circuits). 

The chief contributions of this paper are as follows: {\em(i)} The definition of the concept of a verification functor, between traced symmetric monoidal categories and the category $\Hoa$ (Section \ref{abs-hoare}). {\em(ii)} An abstraction of Hoare triples in terms of verification functors (Definition \ref{abs-triple}). In general, our abstraction of Hoare logic provides a `categorical' recipe for the development of new (automatically) sound and complete Hoare-logic-like rules for any class of systems having the underlying structure of a traced symmetric monoidal category. Moreover, Hoare logic notions such as expressiveness conditions, relative completeness \cite{Cook:1978} and loop invariants, have a clear cut correspondence to some of our abstract notions. {\em(iii)} Sound and complete rules for our abstract notion of Hoare triples, over a fixed verification functor (Theorem \ref{abs-soundness}). {\em(iv)} Four concrete instances of our abstraction, namely: partial correctness of while programs, partial correctness of pointer programs (separation logic), run-time analysis of while programs, and stream circuits (Section \ref{flowchart}). In Section \ref{conclusion}, we discuss the link  between our work and other abstractions of Hoare logic.

\section{Systems in the Abstract}
\label{section:TSMC}
\label{system}

\begin{figure*}[h]
\begin{center}
\setlength{\unitlength}{8mm}
\begin{picture}(11.0,3.0)
	\thicklines
	\put(0.0,2.5){\vector(1,0){1.0}}
	\put(1.6,2.4){$x := 1$}
	\put(1.0,2.1){\framebox(2.5,0.8)}
	\put(3.5,2.5){\vector(1,0){5.2}}
	\put(0.0,1.0){\vector(1,0){1.0}}
	\put(1.5,0.9){$x := x y$}
	\put(1.0,0.6){\framebox(2.5,0.8)}
	\put(3.5,1.0){\vector(1,0){1.0}}
	\put(4.8,0.9){$y := y-1$}
	\put(4.5,0.6){\framebox(2.5,0.8)}
	\put(7.0,1.0){\line(1,0){1.0}}
	\put(8.0,1.0){\vector(0,1){1.5}}
	\put(8.5,3.0){$y \geq 1$}
	\put(9.0,2.5){\circle{0.6}}
	\put(9.3,2.5){\vector(1,0){1.0}}
	\put(9.0,2.2){\line(0,-1){1.3}}
	\put(9.0,0.9){\vector(1,0){1.3}}
	\put(9.9,2.8){(\emph{no})}
	\put(9.9,1.2){(\emph{yes})}
	\put(10.3,0.9){\line(0,-1){0.9}}
	\put(10.3,0.0){\line(-1,0){10.3}}
	\put(0.0,0.0){\line(0,1){1.0}}
\end{picture}
\end{center}
\caption{Example of feedback for flowcharts}
\label{factorial}
\end{figure*}
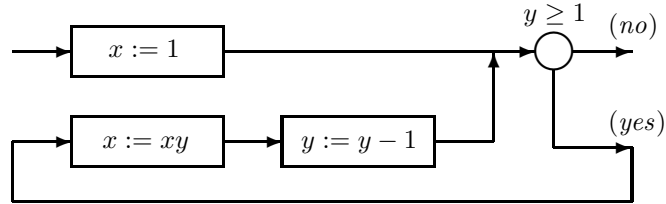

In this section we describe an abstraction of ``system" to be used in our abstraction of the Floyd-Hoare logic \cite{Floyd:1967,Hoare:1969}. 
Consider first the case of flowcharts, or programming languages in general. The essential construct in this case is the unbounded iteration. For instance, in the Figure \ref{factorial} we have depicted a flowchart where the factorial of $y$ is calculated on the variable $x$ via an iterative process. As suggested in the picture, the iteration corresponds to a feedback loop from the (\emph{yes}) branch of the boolean test, to the next step in the computation of the factorial. Note that the code inside the loop ($x := xy; y := y-1$) can be seen as a local process which at each iteration is gearing the computation to achieve the desired result -- the computation of the factorial of $y$.

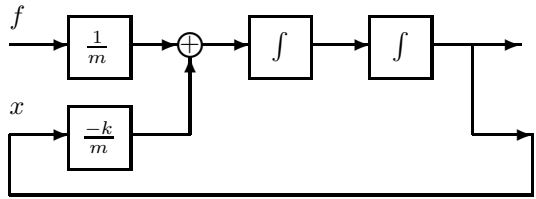
\begin{figure*}[h]
\begin{center}
\setlength{\unitlength}{8mm}
\begin{picture}(9,3.4)
	\thicklines
	\put(0.3,2.9){$f$}
	\put(0.3,2.55){\vector(1,0){1.0}}
	\put(1.55,2.4){$\frac{1}{m}$}
	\put(1.3,2.0){\framebox(1,1)}
	\put(2.3,2.55){\vector(1,0){0.8}}
	\put(3.13,2.44){$+$}
	\put(3.3,2.55){\circle{0.4}}
	\put(0.3,1.4){$x$}
	\put(0.3,1.05){\vector(1,0){1.0}}
	\put(1.5,0.9){$\frac{-k}{m}$}
	\put(1.3,0.5){\framebox(1,1)}
	\put(2.3,1.05){\line(1,0){1.0}}
	\put(3.3,1.05){\vector(0,1){1.3}}
	\put(3.5,2.55){\vector(1,0){0.8}}
	\put(4.65,2.4){$\int$}
	\put(4.3,2.0){\framebox(1,1)}
	\put(5.3,2.55){\vector(1,0){1.0}}
	\put(6.65,2.4){$\int$}
	\put(6.3,2.0){\framebox(1,1)}
	\put(7.3,2.55){\vector(1,0){1.5}}
	\put(8.0,2.55){\line(0,-1){1.5}}
	\put(8.0,1.05){\vector(1,0){1.0}}
	\put(9.0,1.05){\line(0,-1){1.0}}
	\put(9.0,0.05){\line(-1,0){8.7}}
	\put(0.3,0.05){\line(0,1){1.0}}
\end{picture}
\caption{Example of feedback for dynamic systems}
\label{openDiag}
\end{center}
\end{figure*}

Similarly, in the case of continuous (dynamical) systems, the feedback operation is essential for allowing the correction of minor errors and achieving stability. In Figure \ref{openDiag} we depict the control law diagram for a system consisting of a cart (mass $m$) attached to a wall via a spring (constant $k$) being pulled away from the wall via a force $f$.  This diagrammatic expression of the differential equation $m\ddot{x} +kx - f = 0$ might serve, for example, as the design for an analogue computer to simulate the mechanical system. Similarly to the example of while programs described above, in this case the spring acts inside the feedback loop ($-k/m$) as a controller, trying to keep the cart within a particular region.

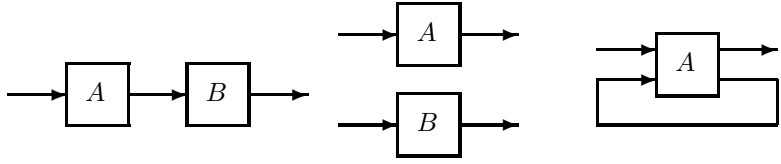
\begin{figure*}[t]
\setlength{\unitlength}{8mm}
\begin{picture}(14,2.5)
	\thicklines
	\put(0.5,1.0){\vector(1,0){1.0}}
	\put(1.5,0.5){\framebox(1,1)}
	\put(1.8,0.9){$A$}
	\put(2.5,1.0){\vector(1,0){1.0}}
	\put(3.5,0.5){\framebox(1,1)}
	\put(3.8,0.9){$B$}
	\put(4.5,1.0){\vector(1,0){1.0}}
	\put(7.3,1.9){$A$}
	\put(6.0,2.0){\vector(1,0){1.0}}
	\put(7.0,1.5){\framebox(1,1)}
	\put(8.0,2.0){\vector(1,0){1.0}}
	\put(7.3,0.4){$B$}
	\put(6.0,0.5){\vector(1,0){1.0}}
	\put(7.0,0.0){\framebox(1,1)}
	\put(8.0,0.5){\vector(1,0){1.0}}
	\put(11.6,1.4){$A$}
	\put(10.3,1.75){\vector(1,0){1.0}}
	\put(10.3,1.25){\vector(1,0){1.0}}
	\put(11.3,1.0){\framebox(1,1)}
	\put(12.3,1.75){\vector(1,0){1.0}}
	\put(12.3,1.25){\line(1,0){1.0}}
	\put(10.3,0.5){\line(1,0){3.0}}
	\put(10.3,0.5){\line(0,1){0.75}}
	\put(13.3,0.5){\line(0,1){0.75}}
\end{picture}
\vspace{2mm}
\caption{Block diagram constructors: sequential composition $\Seq{A}{B}$, parallel composition $\Par(A, B)$, and feedback $\Fb(A)$, respectively.}
\label{syntax}
\end{figure*}

As exemplified above, systems of very different kinds are in general often described via \emph{block diagrams}, with a fixed block-diagram language $\Lang$. Systems in $\Lang$ are built from a set $\bLang$ of basic building blocks via sequential and parallel composition, and feedback (see Figure \ref{syntax}). Let us adopt the convention of using $A, B, C$ for the syntactic block diagrams.

On closer inspection, one will notice that the main properties of the sequential composition are captured by the basic structure of a category\footnote{For the rest of this article we will assume some basic knowledge of category theory. For a readable introduction see \cite{Maclane:1998}. Given a category $\catS$, we will denote its objects by $\catSo$ and its morphisms by $\catSm$. Composition between two morphisms will be denoted as usual by $(g \circ f) : X \to Z$, if $f : X \to Y$ and $g : Y \to Z$.}. In order to capture also the intrinsic properties of the parallel composition $\Par(A,B)$ and feedback $\Fb(A)$, a particular class of categories has been singled out, so-called \emph{traced symmetric monoidal categories}.

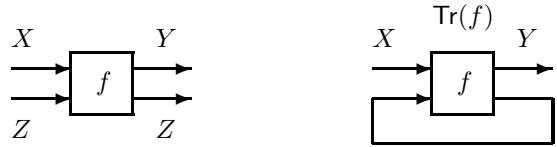
\begin{figure*}[h]
\begin{center}
\setlength{\unitlength}{8mm}
\begin{picture}(10.0,2.5)
	\thicklines
	\put(1.7,0.9){$f$}
	\put(0.3,1.6){$X$}
	\put(0.3,1.25){\vector(1,0){1.0}}
	\put(0.3,0.1){$Z$}
	\put(0.3,0.75){\vector(1,0){1.0}}
	\put(1.3,0.5){\framebox(1,1)}
	\put(2.7,1.6){$Y$}
	\put(2.3,1.25){\vector(1,0){1.0}}
	\put(2.7,0.1){$Z$}
	\put(2.3,0.75){\vector(1,0){1.0}}
	\put(7.3,2.0){$\Tr(f)$}
	\put(7.7,0.9){$f$}
	\put(6.3,1.6){$X$}
	\put(6.3,1.25){\vector(1,0){1.0}}
	\put(6.3,0.75){\vector(1,0){1.0}}
	\put(7.3,0.5){\framebox(1,1)}
	\put(8.7,1.6){$Y$}
	\put(8.3,1.25){\vector(1,0){1.0}}
	\put(8.3,0.75){\line(1,0){1.0}}
	\put(6.3,0.0){\line(1,0){3.0}}
	\put(6.3,0.0){\line(0,1){0.75}}
	\put(9.3,0.0){\line(0,1){0.75}}
\end{picture}
\end{center}
\caption{Trace diagrammatically}
\label{trace}
\end{figure*}

Recall that a pair comprising a category $\catS$ and a covariant bifunctor $\otimes$ is called a \emph{monoidal category} if for some particular object of $\catS$ (the identity element of the monoid) $\otimes$ satisfies the monoidal axioms of associativity and identity (for details, see \cite[chapter 11]{Maclane:1998}). A monoidal category is called symmetric if there exists a family of natural isomorphisms $c_{X, Y} : X \otimes Y \to Y \otimes X$ satisfying the two braiding axioms plus the symmetry axiom $c_{X, Y} \circ c_{Y, X} = \id_{X \otimes Y}$. In \cite{Joyal:1996}, the notion of \emph{traced symmetric monoidal category} is introduced\footnote{In fact, \cite{Joyal:1996} introduces the theory of traces for a more general class of monoidal categories, so-called balanced monoidal categories, of which symmetric monoidal categories are a special case.}, for short \emph{TMC}. A symmetric monoidal category is traced if for any morphism $f : X \otimes Z \to Y \otimes Z$ there exists a morphism $\Tr_{X, Y}^Z(f) : X \to Y$ satisfying the trace axioms (see \cite{Joyal:1996} for details). We will normally omit the decoration in $\Tr_{X, Y}^Z$ whenever it is clear over which objects the trace is being applied. Morphisms of a TMC can be represented diagrammatically as input-output boxes, so that, if $f : X \otimes Z \to Y \otimes Z$ then $\Tr(f) : X \to Y$ corresponds to a \emph{feedback} over the `wire' $Z$, as shown in Figure \ref{trace}. In Sections \ref{flowchart} and \ref{network} we will give the formal definition of TMCs based on disjoint union (flowcharts) and cartesian product (networks). Before that we define the necessary concepts for our abstract Hoare logic system.


In our work, we are interested in traced monoidal categories which arise as the semantics of a syntactic block diagram language $\Lang$. More precisely, for each block diagram language $\Lang$ we will consider (possibly different) semantic mappings $\sm{\cdot}$ into particular traced monoidal categories $\catS$. We will use $f, g, h$ for the semantic functions denoted by these block diagrams. We assume that the semantic mapping is compositional, i.e.
\begin{eqnarray*}
\sm{\Seq{A}{B}} &=& \sm{B} \circ \sm{A} \\
\sm{\Par(A, B)} &=& \sm{A} \otimes \sm{B} \\
\sm{\Fb(A)}     &=& \Tr_{\catS}(\sm{A}).
\end{eqnarray*}
Note that the objects and morphisms in the range of such a semantic mapping form a sub-TMC of $\catS$ and we can view $\Lang$ as a typed language in which the types identify hom-sets in this sub-TMC and where the constructors combine the types of their operands as follows:
\[
\begin{array}{lcl}
(\Seq{\cdot\,}{\cdot}) &\quad : \quad& (\alpha \to \beta) \times (\beta \to \gamma) \Rightarrow \alpha \to \gamma \\[2mm]
\Par(\cdot\,, \cdot) &:& (\alpha \to \beta) \times (\gamma \to \delta) \Rightarrow (\alpha \otimes \gamma) \to (\beta \otimes \delta) \\[2mm]
\Fb(\cdot)        &:& (\alpha \otimes \gamma) \to (\beta \otimes \gamma) \Rightarrow \alpha \to \beta.
\end{array}
\]

\section{Abstract Hoare Logic}
\label{abs-hoare}

Recall that a pre-order is a pair $(X, \leq)$ consisting of a set $X$ and a binary relation $\leq$ on $X$ which is reflexive and transitive. For instance, the set of formulas of a fixed first-order theory under the consequence relation (i.e. $A \vdash B$) forms a pre-order, or any set of sets under subset inclusion $\subseteq$.

Let $A$ be a while program, $P, Q$ be assertions (formulas) over the program variables of $A$, and $\hTriple{P}{A}{Q}$ denote the usual Hoare triple for while programs. The only property of the binary relation $\hTriple{\cdot}{A}{\cdot}$ which does not depend on the structure of $A$, and hence is intrinsic to the Hoare triples themselves, is the monotonicity stated in the \emph{consequence rule}:
\begin{itemize}
	\item[] if $\hTriple{P}{A}{Q}$ holds and $P' \to P$ and $Q \to Q'$ then $\hTriple{P'}{A}{Q'}$ also holds.
\end{itemize}
The following definition captures this basic property of assertions and Hoare triples. When $r$ is a binary relation we write $\rel{x}{r}{y}$ as a shorthand for $\pair{x}{y} \in r$.

\begin{definition}[Hoare category] A relation $r : X \times Y$ between two pre-ordered sets $X, Y$ is called \emph{monotone} if $\rel{P}{r}{Q}$ and $P' \iord_X P$ and $Q \iord_Y Q'$ implies $\rel{P'}{r}{Q'}$. Let $\Hoa$ denote the category of pre-ordered sets and monotone relations. We call $\Hoa$ the \emph{Hoare category}.
\end{definition}

It is easy to see that $\Hoa$ can be considered a symmetric monoidal category, with the monoidal operation as cartesian product, since the cartesian product of two pre-ordered sets $X, Y$ forms again a pre-ordered set with the order on $X \times Y$ defined coordinatewise, i.e., $\pair{x}{y} \iord \pair{x'}{y'}$ iff $x \iord x'$ and $y \iord y'$. It is also easily verified that the trace based on cartesian product (see Section \ref{network}) gives a trace on $\Hoa$, making it into a TMC. Given a block diagram language $\Lang$, with its respective trace monoidal semantics, we define an abstract notion of Hoare logic using certain functors from the semantic domains into $\Hoa$:


\begin{definition}[(Verification functor)] Let $\Lang$ be a fixed block diagram language with semantics $\sm{\cdot} : \Lang \to \catS$. A mapping  $H : \catS \to \Hoa$ is called a \emph{verification functor} if it behaves like a strict traced monoidal functor on the image (a sub-TCM of $\catS$) of the semantic mapping $\sm{\cdot}$ (see Figure \ref{categories}).
\end{definition}

\begin{figure*}[t]
\begin{center}
\setlength{\unitlength}{8mm}
\begin{picture}(8,2.0)
	\thicklines
	\put(1.0,1.0){\oval(1.5,2.0)}
	\put(0.9,0.9){$\Lang$}
	\put(2.2,1.3){$\sm{\cdot}$}
	\put(2.0,1.0){\vector(1,0){1.0}}
	\put(4.0,1.0){\oval(1.5,2.0)}
	\put(3.85,0.9){$\catS$}
	\put(5.2,1.3){$H$}
	\put(5.0,1.0){\vector(1,0){1.0}}
	\put(7.0,1.0){\oval(1.5,2.0)}
	\put(6.8,0.9){$\Hoa$}
\end{picture}
\end{center}
\caption{Block diagram language $\Lang$, tr. mon. sematic $\catS$, and Hoare category $\Hoa$.}
\label{categories}
\end{figure*}

Let a verification functor $H : \catS \to \Hoa$ be fixed. Each object $X \in \catSo$ corresponds to a pre-ordered set $H(X)$, and each morphism $f$ in $\catS$ corresponds to a monotone relation $H(f)$. Under the assumption that the semantic mapping $\sm{\cdot}$ is compositional, the verification functor condition is equivalent to:
\begin{itemize}
	\item[(SC1)] $H(\sm{\Seq{A}{B}}) = H(\sm{B}) \circ H(\sm{A})$ \\[-2mm]
	\item[(SC2)] $H(\sm{\Par(A, B)}) = H(\sm{A}) \times H(\sm{B})$ \\[-2mm]
	\item[(SC3)] $H(\sm{\Fb(A)}) = \Tr_{\Hoa}(H(\sm{A}))$
\end{itemize}
for all block diagrams $A, B$ in the block diagram language $\Lang$. Relational equality will be needed for soundness and completeness of the corresponding Hoare logic. As we will see, if only soundness is required only relational inclusion is needed, i.e.
\begin{itemize}
	\item[(S1)] $H(\sm{\Seq{A}{B}}) \supseteq H(\sm{B}) \circ H(\sm{A})$ \\[-2mm]
	\item[(S2)] $H(\sm{\Par(A, B)}) \supseteq H(\sm{A}) \times H(\sm{B})$ \\[-2mm]
	\item[(S3)] $H(\sm{\Fb(A)}) \supseteq \Tr_{\Hoa}(H(\sm{A}))$
\end{itemize}

\begin{definition}[Abstract Hoare Triples] \label{abs-triple} Let $A$ be a concrete block diagram, whose meaning is a morphism $\sm{A} : X \to Y$ in $\catS$. Moreover, let $P \in H(X)$ and $Q \in H(Y)$. Define \emph{abstract Hoare triples} as
\begin{equation}
\hTriple{P}{A}{Q} \;\; \pdefin \;\; \rel{P}{H(\sm{A})}{Q}
\end{equation}
\end{definition}

Although we use the same notation as the standard Hoare triple, it should be noted that the meaning of our abstract Hoare triple can only be given once the verification functor $H$ is fixed. The usual Hoare logic meaning of \emph{if $P$ holds before the execution of $A$ then, if $A$ terminates, $Q$ holds afterwards} will be one of the special cases of our general theory. See Sections \ref{flowchart} and \ref{network} for other meanings of $\hTriple{P}{f}{Q}$.

Let a specific block diagram language be fixed, and let $\catS$ be the traced monoidal category denoting the programs of the given language. Moreover, let $H : \catS \to \Hoa$ be a fixed verification functor. We will denote by $\Hsystem{\Lang}{\catS}{H}$ the set of rules shown in Figure \ref{rules}, where the side conditions are:
\begin{itemize}
	\item[($\dagger$)] $A \in \bLang$ and $\rel{P}{H(\sm{A})}{Q}$ \\[-2mm]
	\item[($\ddagger$)] $P' \iord P$ and $Q \iord Q'$
\end{itemize}

The formal system $\Hsystem{\Lang}{\catS}{H}$ should be viewed as a \emph{syntactic} axiomatisation\footnote{Note that assertions $P, Q, R$ could potentially be semantic objects (e.g. sets), which is harmless since the only structure required from assertions is the ability to form a pair $\pair{P}{Q}$ out of two assertions $P$ and $Q$. Moreover, the fact that logical axioms have a semantic side condition $(\dagger)$ is not too restrictive since this is only assumed for the basic block diagrams, and these are normally manageable computationally.} of the ternary relation $\hTriple{P}{A}{Q}$. The verification functor $H$ gives the \emph{semantics} of the Hoare triples (and rules). By soundness and completeness of the system $\Hsystem{\Lang}{\catS}{H}$ we mean that syntax corresponds precisely to semantics, i.e. a syntactic Hoare triple $\hTriple{P}{A}{Q}$ is provable in $\Hsystem{\Lang}{\catS}{H}$ if and only if $\rel{P}{H(\sm{A})}{Q}$ is true in $\Hoa$.

\begin{figure*}[t]
\[
\begin{array}{|cc|}
\hline
& \\
\quad \begin{prooftree}
\justifies
\hTriple{P}{A}{Q}
\using{({\sf Ax}) \; (\dagger)}
\end{prooftree} \quad
&
\quad \begin{prooftree}
	\hTriple{P}{A}{Q}
	\justifies
	\hTriple{P'}{A}{Q'}
	\using{({\sf Con}) \, (\ddagger)}
\end{prooftree} \quad \\
& \\
\multicolumn{2}{|c|}{
\begin{prooftree}
	\hTriple{P}{A}{Q} \quad \hTriple{R}{B}{S}
	\justifies
	\hTriple{\pair{P}{R}}{\Par(A, B)}{\pair{Q}{S}}
	\using{(\Par)}
\end{prooftree}} \\[4mm]
& \\
\quad \begin{prooftree}
	\hTriple{P}{A}{Q} \quad \hTriple{Q}{B}{R}
	\justifies
	\hTriple{P}{\Seq{A}{B}}{R}
	\using{({\sf Seq})}
\end{prooftree} \quad
&
\begin{prooftree}
	\hTriple{\pair{P}{Q}}{A}{\pair{R}{Q}}
	\justifies
	\hTriple{P}{\Fb(A)}{R}
	\using{(\Fb)}
\end{prooftree} \quad \\
& \\
\hline
\end{array}
\]
\caption{The system $\Hsystem{\Lang}{\sm{\cdot}}{H}$ with $\sm{\cdot} : \Lang \to \catS$ and $H : \catS \to \Hoa$.}
\label{rules}
\end{figure*}

\begin{theorem}[Soundness and completeness] \label{abs-soundness} The system $\Hsystem{\Lang}{\sm{\cdot}}{H}$ is sound and complete.
\end{theorem}
{\bf Proof.} Soundness is trivially true for the axioms. The consequence rule is sound by the monotonicity of the relation $H(\sm{A})$. Soundness of the composition rule also uses the fact that $H$ respects composition, i.e.
\[ \rel{P}{H(\sm{\Seq{A}{B}})}{R} \quad \Leftrightarrow \quad \rel{P}{H(\sm{B}) \circ H(\sm{A})}{R}. \]
Soundness of the cartesian product rule uses that the functor $H$ is monoidal, i.e.
\[ \rel{\pair{P}{Q}}{H(\sm{A \otimes B})}{\pair{R}{S}} \quad \Leftrightarrow \quad \rel{\pair{P}{Q}}{H(\sm{A}) \times H(\sm{B})}{\pair{R}{S}}. \]
For the soundness of the trace rule, assume $\rel{\pair{P}{Q}}{H(\sm{A})}{\pair{R}{Q}}$. By the definition of the trace on $\Hoa$ we have $\rel{P}{\Tr(H(\sm{A}))}{R}$. Finally, by the fact that $H$ is traced we have $\rel{P}{H(\sm{\Fb(A)})}{R}$. We argue now about completeness. It is easy to verify that all true statements of the form $\hTriple{P}{A}{Q}$, for basic diagrams $A$, are provable. It remains to show that if $\hTriple{P}{A}{Q}$ is true, for an arbitrary block diagram $A$, then there exists a premise of the corresponding rule (depending on the structure of $A$) which is also true. If $\hTriple{P}{\Seq{A}{B}}{R}$ is true then, since $H$ respects composition, there must exists a $Q$ such that both
\[ \rel{P}{H(\sm{A})}{Q} \quad \mbox{and} \quad \rel{Q}{H(\sm{B})}{R} \]
are true. If $\hTriple{\pair{P}{R}}{A \otimes B}{\pair{Q}{S}}$ is true then so it is $\hTriple{P}{A}{Q}$ and $\hTriple{R}{B}{S}$, by the fact that $H$ is monoidal. Finally, if $\hTriple{P}{\Fb(A)}{R}$ is true, then so is
\[ \rel{\pair{P}{Q}}{H(\sm{A})}{\pair{R}{Q}}, \]
for some $Q$ by the definition of a verification functor. \eProof

The abstract proof of soundness and completeness presented above is both short and simple, using only the assumption of a verification functor and basic properties of the category $\Hoa$. As we will see, the laborious work is pushed into showing that $H$ is a verification functor. That will be clear in Section \ref{while}, where we build a verification functor appropriate for while programs, using Cook's expressiveness condition \cite{Cook:1978}. 



\section{Verification Functors for Flowcharts}
\label{flowchart}

The traced monoidal categories that correspond to classical Hoare logics for imperative programs are based on the category of sets and relations with the monoidal structure arising from the disjoint union operator $\uplus$. We will write $\inj_i : X_i \to X_0 \uplus X_1$ for the injections of the summands into a disjoint union. In this TMC, the trace of a relation $f : X \uplus Z \to Y \uplus Z$ is then defined as follows:
\eqleft{\rel{x}{\Tr(f)}{y} \;\pdefin\; \exists z_0 \ldots z_n (\rel{\linj{x}}{f}{\rinj{z_0}} \wedge \ldots \rel{\rinj{z_i}}{f}{\rinj{z_{i+1}}} \ldots \wedge \rel{\rinj{z_n}}{f}{\linj{y}})}
This trace based on disjoint union lets us view feedback loops as representing iterative processes. 

First we will consider systems (block diagrams) where disjoint union is used for the monoidal structure in the semantic domains. These are normally called ``flowcharts", and two instances of these are (refinements of) while programs and pointer programs. The instantiations will produce the original Hoare logic \cite{Hoare:1969}, separation logic and a new Hoare logic for running time analysis. 

\begin{figure*}[t]
\setlength{\unitlength}{8mm}
\begin{picture}(14,4)
	\thicklines
	\put(3.5,3.5){Identity}
	\put(3.5,3.0){\vector(1,0){1.5}}
	\put(10.0,3.5){Join}
	\put(8.5,3.5){\line(2,-1){1.0}}
	\put(8.5,2.5){\line(2,1){1.0}}
	\put(9.5,3.0){\vector(1,0){1.0}}
	\put(1.3,1.5){Assignment}
	\put(0.5,0.5){\vector(1,0){1.0}}
	\put(1.5,0.0){\framebox(1.5,1)}
	\put(1.7,0.4){$x \assign t$}
	\put(3.0,0.5){\vector(1,0){1.0}}
	\put(6.5,1.5){Twist}
	\put(6.0,1.0){\vector(2,-1){2.0}}
	\put(6.,0.0){\vector(2,1){2.0}}
	\put(10.5,1.5){Conditional}
	\put(10.0,0.5){\vector(1,0){1.0}}
	\put(11.5,0.5){\circle{1.0}}
	\put(11.4,0.35){$b$}
	\put(11.95,0.2){\vector(4,-1){1.0}}
	\put(11.95,0.8){\vector(4,1){1.0}}	
\end{picture}
\vspace{2mm}
\caption{Basic flowcharts.}
\label{basicFlow}
\end{figure*}
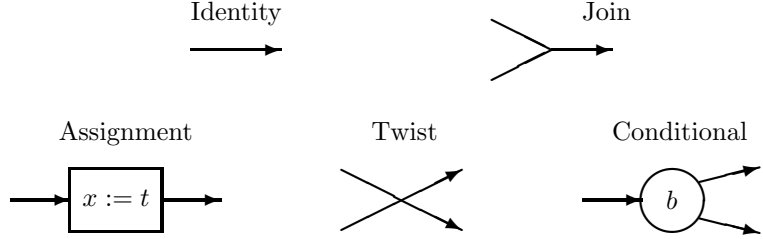

\subsection{The programming language: flowcharts}
\label{while-prog}

\begin{figure*}[t]
\setlength{\unitlength}{8mm}
\begin{picture}(14,2.8)
	\thicklines
	\put(0.5,1.5){De-reference}
	\put(0.0,0.5){\vector(1,0){1.0}}
	\put(1.0,0.0){\framebox(1.5,1)}
	\put(1.15,0.4){$x \assign [t]$}
	\put(2.5,0.5){\vector(1,0){1.0}}
	\put(4.3,2.5){Reference}
	\put(3.5,1.5){\vector(1,0){1.0}}
	\put(4.5,1.0){\framebox(1.5,1)}
	\put(4.65,1.4){$[t] \assign s$}
	\put(6.0,1.5){\vector(1,0){1.0}}
	\put(8.1,1.5){New cells}
	\put(6.9,0.5){\vector(1,0){1.0}}
	\put(7.9,0.0){\framebox(2.3,1)}
	\put(8.05,0.4){$x \assign {\sf new}(\vec{t})$}
	\put(10.2,0.5){\vector(1,0){1.0}}
	\put(11.8,2.5){Dispose}
	\put(11.0,1.5){\vector(1,0){1.0}}
	\put(12.0,1.0){\framebox(1.4,1)}
	\put(12.15,1.4){${\sf disp}(t)$}
	\put(13.4,1.5){\vector(1,0){1.0}}
\end{picture}
\vspace{2mm}
\caption{Basic pointer programs.}
\label{basicPointer}
\end{figure*}
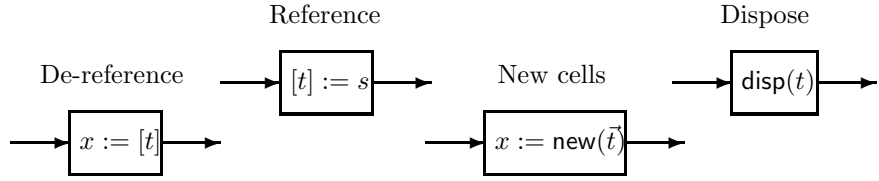

In the case of simple diagrammatic while programs the basic constructs of our language $\fLang$ are shown in Figure \ref{basicFlow}. These can be put together via sequential composition, parallel composition and feedback, as shown in Figure \ref{syntax}. We will consider an extension $\pLang$ of this programming language with pointers by adding the basic pointer programs shown in Figure \ref{basicPointer}.

\subsection{The semantics}

The programs (flowcharts) which can be built as described in Section \ref{while-prog} can be given a denotation as follows. Let $\Store$ be the set mappings $\rho : \Var \to \ZZ$ assigning an integer value to each of the program variables $\Var = \{x, y, \ldots\}$. Let the objects of $\fcatS$ be the set (of sets) containing the empty set $\emptyset$ and $\Store$, and closed under \emph{disjoint union}, i.e. $\{\emptyset, \Store, \Store \uplus \Store, \ldots\}$. Consider also the following family of functions between the objects of $\fcatS$: 
%
\begin{itemize}
	\item[-] {\it Skip}, $\nops : \Store \to \Store$
	\item[] \quad $\nops(\rho) \fdefin \rho$ \\[-2mm]
	\item[-] {\it Assignment}, $(x \assign t) : \Store \to \Store$
	\item[] \quad $(x \assign t)(\rho) \fdefin (\rho)[t_\rho/x]$ \\[-2mm]
	\item[-] {\it Joining}, $\Delta : \Store \uplus \Store \to \Store$
	\item[] \quad $\Delta({\sf inj}_i(\rho)) \fdefin \rho$ \\[-2mm]
	\item[-] {\it Twist}, $c : \Store \uplus \Store \to \Store \uplus \Store$
	\item[] \quad $\twist(\pair{v}{\rho}) \fdefin \pair{\overline{v}}{\rho}$ \\[-2mm]
	\item[-] {\it Forking}, $\cond{b} : \Store \to \Store \uplus \Store$
	\item[] \quad $\cond{b}(\rho) \fdefin
	\left\{
		\begin{array}{ll}
		\linj(\rho) \quad & {\sf if} \; \neg b_\rho \\
		\rinj(\rho) & {\sf otherwise}
		\end{array}
	\right.$ 
\end{itemize}
The conditional forking ($\cond{b}$) and the assignment ($x \assign t$) are parametrised by functions $b$ and $t$ (intuitively, expressions) from $\Store$ to the boolean lattice $\BB$ and $\ZZ$, respectively, so that $b_\rho$ and $t_\rho$ denote their value on a given store $\rho$. We use $\linj$ and $\rinj$ for left and right injections into $\Store \uplus \Store$. 

We then close the set of basic functions in $\fcatS$ under sequential composition of functions, disjoint union of functions and the standard trace for disjoint union, to form the set of partial functions which are the morphisms of $\fcatS$. It is easy to see that $\fcatS$ forms a TMC.

In the case of pointer programs, we consider an adaptation $\fcatS$ of the above TMC to a category of programs that manipulate both stores and heaps, which we will refer to as the traced symmetric monoidal category of \emph{pointer programs} $\pcatS$.  Let $\State \fdefin \Store \times \Heap$, where $\Store$ is as above and $\Heap$ is the set of partial functions (from $\NN$ to $\ZZ$) with finite domain. Define also for any set $X$ a new set $X_{\sf a}$ as $X \cup \{\abort\}$. We view the elements of the heap as pairs consisting of a function $h : \NN \to \ZZ$ and a finite set $d \in {\cal P}_{{\sf fin}}(\NN)$ describing the valid domain of $h$. We then form the set (of sets) of objects of $\pcatS$ as the set containing the empty set $\emptyset$ and $\State_{\sf a}$, and closed under \emph{disjoint union} but maintaining $\abort$ as a form of bottom element, i.e. $\{\emptyset, \State_{\sf a}, (\State \uplus \State)_{\sf a}, \ldots\}$. It is easy to check that $\pcatS$ is also a TMC. Each of the basic functions of $\fcatS$ can be lifted to the extended type structure, by simply ignoring the `heap' component, or by propagating $\abort$ when receiving an $\abort$ as input. The set of basic functions of $\pcatS$ is an extension of the set of (the lifting of the) basic functions of $\fcatS$ with the following family of functions:
%
\begin{itemize}
	\item {\it Look up}, $(x \assign [t]) : \State_{\sf a} \to \State_{\sf a}$
	\item[] \quad $(x \assign [t])(\rho, h, d) \fdefin \left\{
		\begin{array}{ll}
		(\rho[h(t_\rho)/x], h, d) \quad & t_\rho \in d \\
		\abort \quad & {\sf otherwise}
		\end{array}
	\right.$ \\
	\item {\it Mutation}, $([t] \assign s) : \State_{\sf a} \to \State_{\sf a}$
	\item[] \quad $([t] \assign s)(\rho, h, d) \fdefin \left\{
		\begin{array}{ll}
		(\rho, h[t_\rho \mapsto s_\rho], d) \quad & t_\rho \in d \\
		\abort \quad & {\sf otherwise}
		\end{array}
	\right.$ \\
	\item {\it Allocation}, $x \assign {\sf new}(\vec{t}) : \State_{\sf a} \to \State_{\sf a}$
	\item[] \quad $(x \assign {\sf new}(\vec{t}))(\rho, h, d) \fdefin (\rho[x \mapsto i], h[i+j \mapsto t_j], d \uplus \{i, \ldots, i + n\})$ \\[-2mm]
	\item {\it Deallocation}, ${\sf disp}(t) : \State_{\sf a} \to \State_{\sf a}$
	\item[] \quad $({\sf disp}(t))(\rho, h, d) \fdefin \left\{
		\begin{array}{ll}
		(\rho, h, d\backslash\{t_\rho\}) \quad & t_\rho \in d \\
		\abort \quad & {\sf otherwise}
		\end{array}
	\right.$
\end{itemize}

In order to make the allocation functional deterministic, we let $i$ be the address location succeeding the maximum address already defined.

As done in the case of flowcharts above, we can then close the set of basic functions under sequential composition, disjoint union and trace, to form the set of partial functions which are the morphisms of $\pcatS$. We are assuming that the functions are strict with respect to $\abort$, i.e. on the $\abort$ state all programs will return $\abort$. The category $\pcatS \equiv (\pcatS, \uplus, \Tr)$, with the standard trace for disjoint union, forms another example of a TMC with the extra basic morphisms look up, mutation, allocation and deallocation. 

\subsection{Hoare logic for partial correctness}
\label{while}

In this section we present a verification embedding of the TMC of flowcharts $\fcatS$. This will give us soundness and (relative) completeness of Hoare's original verification logic \cite{Hoare:1969} for partial correctness (using forward reasoning).

Let us now define the monoidal functor $\fcH : \fcatS \to \Hoa$. Let a Cook-expressive\footnote{Recall that a logic is Cook-expressive if for any program $f$ and pre-condition $P$ the strongest post-condition of $f$ under $P$ is expressible by a formula in $\Lang$ (cf. \cite{Cook:1978}).} first-order theory be fixed. On the objects $X \in \fcatS$ we let $\fcH(X)$ be a pre-ordered set of formulas. The ordering $P \iord R$ on the elements of $\fcH(X)$ is taken to be $P \to R$ in the fixed theory. We assume that pairs of formulas are also considered formulas, with the connectives defined pointwise. We now define the functor $\fcH$ on the image of the semantic embedding $\sm{\cdot}$. For each flowchart $A$ we let $\fcH(\sm{A})$ be the following monotone relation
\eqleft{\rel{P}{\fcH(\sm{A})}{Q} \,\fdefin\, \sPost(A, P) \to Q}
where $\sPost(A, P)$ is a formula expressing the strongest post-condition of $A$ under $P$. Such formula exists by our assumption that the theory is Cook-expressive. Moreover, note that the denotation of $\rel{}{\fcH(\sm{A})}{}$, since it only depends on $A$ via $\sPost(A, \cdot)$, does not depend on the particular syntax of $A$, but only on the input-output behaviour of $A$, i.e. $\sm{A}$. It is also easy to see what $\sPost$ is for the basic morphisms of $\fcatS$
\eqleft{
\begin{array}{lcl}
	\sPost(\nops, P) & \fdefin & P \\[2mm]
	\sPost(x \assign t, P) & \fdefin & \exists v (P[v/x] \wedge x = t[v/x]) \\[2mm]
	\sPost(\cond{b}, P) & \fdefin & \pair{P \wedge \neg b}{P \wedge b} \\[2mm]
	\sPost(\twist, \pair{P}{R}) & \fdefin & \pair{R}{P} \\[2mm]
	\sPost(\Delta, \pair{P}{R}) & \fdefin & P \vee R
\end{array}}

The functor $H$ is monoidal because a formula $P$ describing a subset of $X_0 \uplus X_1$ can be seen as a pair of formulas $\pair{P_0}{P_1}$ such that each $P_i$ describes a subset of $X_i$, i.e. $\fcH(X \uplus Y)$ is isomorphic to $\fcH(X) \times \fcH(Y)$. Similarly, there is a one-to-one correspondence between  strongest post-condition transformer for a parallel composition of flowcharts $f \uplus g$ and pairs of predicate transformers $\fcH(f) \times \fcH(g)$. 

We argue now in two steps that $\fcH$ is also a verification functor. The main task is to show that $\fcH$ respects the trace structure, i.e.
\[ \rel{P}{\fcH(\Tr(\sm{A}))}{R} \quad \Leftrightarrow \quad \rel{P}{\Tr(\fcH(\sm{A}))}{R}. \]
By the definition of trace on the $\Hoa$ we get
\[ \rel{P}{\fcH(\Tr(\sm{A}))}{R} \quad \Leftrightarrow \quad \exists Q \; (\rel{\pair{P}{Q}}{\fcH(\sm{A})}{\pair{R}{Q}}) \]
and by the definition of $\fcH$ above:
\[ \sPost(\Tr(\sm{A}))(P) \to R \quad \Leftrightarrow \quad \exists Q \; (\sPost(\sm{A})\pair{P}{Q} \to \pair{R}{Q}). \]
The next lemma proves the left to right implication. The implication from right to left is proven in Theorem \ref{while-trace}. 

\begin{figure*}[t]
\begin{center}
\setlength{\unitlength}{6mm}
\begin{picture}(16,3.2)
	\thicklines
	\put(3.0,3){$A'$}
	\put(11.0,3){$\Fb(A')$}
	\put(0.4,1.8){\vector(1,0){1.0}}
	\put(0.4,0.8){\vector(1,0){1.0}}
	\put(1.4,0.3){\framebox(2,2)}
	\put(2.15,1.2){$A$}
	\put(3.4,1.8){\vector(1,0){3.0}}
	\put(3.4,0.8){\vector(1,0){1.0}}
	\put(4.7,1.2){$\cond{\vec{z} = \vec{y}}$}
	\put(4.7,0.8){\circle{0.6}}
	\put(5.0,0.8){\vector(1,0){1.4}}
	\put(4.7,0.5){\line(0,-1){0.5}}
	\put(4.7,0.0){\vector(1,0){1.7}}
	\put(0.0,2.1){$X$}
	\put(0.0,1.1){$Z$}
	\put(6.4,2.1){$Y$}
	\put(6.4,1.1){$Z$}
	\put(6.4,0.2){$Z$}
	\put(9.0,1.8){\vector(1,0){1.0}}
	\put(9.0,0.8){\vector(1,0){1.0}}
	\put(10.0,0.3){\framebox(2,2)}
	\put(10.8,1.2){$A$}
	\put(12.0,1.8){\vector(1,0){3.0}}
	\put(12.0,0.8){\vector(1,0){1.0}}
	\put(13.3,1.2){$\cond{\vec{z} = \vec{y}}$}
	\put(13.3,0.8){\circle{0.6}}
	\put(13.6,0.8){\vector(1,0){1.4}}
	\put(13.3,0.5){\line(0,-1){0.8}}
	\put(13.3,-0.3){\line(-1,0){4.3}}
	\put(9.0,-0.3){\line(0,1){1.1}}
	\put(8.6,2.1){$X$}
	\put(15.0,2.1){$Y$}
	\put(15.0,1.1){$Z$}
\end{picture}
\end{center}
\caption{Cook's construction (forward reasoning)}
\label{cook-trick}
\end{figure*}
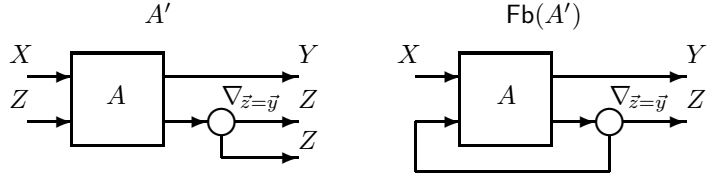

\begin{lemma}[\cite{Cook:1978}] \label{crucial} Let $A$ be a block diagram such that $\sm{A} : X \uplus Z \to Y \uplus Z$. Assume $\vec z$ are all the program variables of $A$. Moreover, let $P \in \fcH(X)$ and $R \in \fcH(Y)$ be fixed formulas. If $\sPost(\Tr(\sm{A}), P) \to R$ then $\sPost(\sm{A}, \pair{P}{Q}) \to \pair{R}{Q}$, for some formula $Q$.
\end{lemma}
{\bf Proof.} We construct a formula $Q$ which is a fixed point for $\sm{A}$ on $P$, i.e. $\sPost(\sm{A}, \pair{P}{Q}) \leftrightarrow \pair{R'}{Q}$, for some formula $R'$. We also argue that
\eqleft{\sPost(\Tr(\sm{A}), P) \leftrightarrow R'.}
By our hypothesis $\sPost(\Tr(\sm{A}), P) \to R$ it then follows that $R'$ implies $R$, as desired. The fixed point $Q$ is essentially the strongest loop invariant, and is constructed as follows. Given the block diagram $A$ we build a new block diagram $A' : X \uplus Z \to Y \uplus Z \uplus Z$ where the internal states of $Z$ can be observed, even after the feedback is applied. Let $A' \fdefin (\id \uplus \cond{\vec{z}=\vec{y}}) \circ A$ (see Figure \ref{cook-trick})
where $\vec{z}$ is the finite sequence of variables mentioned in the description of $A$, and $\vec{y}$ is a fresh tuple of variables of same length. Notice that the program $\Fb^Z_{X, Y \uplus Z}(A')$ behaves almost as $\Fb^Z_{X, Y}(A)$ except that the block diagram $\Fb(A')$ will `terminate' earlier (meaning that the fixed point sequence is shorter) if the state $\vec{z}$ matches $\vec{y}$. Let $\sPost(\sm{\Fb(A')}, P) = \pair{Q_0}{Q_1(\vec{y})}$. 
%
It is easy to see that the formula $Q \equiv \exists \vec{y} \, Q_1(\vec{y})$ characterises the possible internal states $\vec{z}$ in any run of $\Fb(A)$ on an input satisfying $P$.
In other words, $Q$ is the least fixed point for $\fcH(\sm{A})$ on $P$, $\sPost(\sm{A}, \pair{P}{Q}) \leftrightarrow \pair{R'}{Q}$, for some formula $R'$, which implies $\sPost(\Tr(\sm{A}), P) \leftrightarrow R'$. \eProof

\begin{theorem} \label{while-trace} $\fcH : \fcatS \to \Hoa$, as defined above, is a verification functor for morphisms arising from block diagrams in $\fLang$.
\end{theorem}
{\bf Proof.} By Lemma \ref{crucial}, it remains to be shown that whenever
\begin{itemize}
	\item[$(i)$] $\sPost(\sm{A}, \pair{P}{Q}) \to \pair{R}{Q}$,
\end{itemize}
for a formula $Q$, then $\sPost(\Tr(\sm{A}), P) \to R$. Assume $(i)$ and $\sPost(\Tr(\sm{A}), P)(\rho)$, for some store value $\rho$. We must show $R(\rho)$. By the definition of the strongest post-condition there exists a sequence of stores $\rho', \rho_0, \ldots, \rho_n$ such that $P(\rho')$ and
\eqleft{\rel{\pair{0}{\rho'}}{\sm{A}}{\pair{1}{\rho_0}}, \ldots, \rel{\pair{1}{\rho_k}}{\sm{A}}{\pair{1}{\rho_{k+1}}}, \ldots, \rel{\pair{1}{\rho_n}}{\sm{A}}{\pair{0}{\rho}}.}
By a simple induction, using the assumption $(i)$, we get that all $\rho_k$ satisfy $Q$ and that $\rho$ satisfies $R$, as desired. \eProof
 
Given the semantic embedding $\sm{\cdot} : \fLang \to \fcatS$, the system $\Hsystem{\fLang}{\sm{\cdot}}{\fcH}$ obtained via our embedding $\fcH$ is a refinement of the system given by Hoare \citeyear{Hoare:1969}, i.e. Hoare's rules are derivable from ours. See, for instance, the case of the while loop rule in Figure \ref{while-derivation}, given that a while loop $\while_b(A)$ can be represented in $\fcatS$ as $\Fb(\Seq{\Seq{\Delta}{\cond{b}}}{(\nops \uplus A)})$. Moreover, the soundness and (relative) completeness of the Hoare logic rules for while programs follow easily from Theorems \ref{abs-soundness} and \ref{while-trace}.

\begin{figure*}[h]
\[ 
\begin{prooftree}
\[
	\[
		\[
			(\Delta \in \bLang_0)
			\justifies
			\hTriple{\pair{P}{P}}{\Delta}{P}
		\]
		\[
			\[
				(\cond{b} \in \bLang_0)
				\justifies
				\hTriple{P}{\cond{b}}{\pair{P \wedge \neg b}{P \wedge b}}
			\]
			\[
				\[
					(\nops \in \bLang_0)
					\justifies
					\hTriple{P \wedge \neg b}{\nops}{P \wedge \neg b}
				\]
				{\bf \hTriple{P \wedge b}{A}{P}}
				\justifies
				\hTriple{\pair{P \wedge \neg b}{P \wedge b}}{\Par(\nops, A)}{\pair{P \wedge \neg b}{P}}
			\]
			\justifies
			\hTriple{P}{\Seq{\cond{b}}{\Par(\nops, A)}}{\pair{P \wedge \neg b}{P}}
		\]
		\justifies
		\hTriple{\pair{P}{P}}{\Seq{\Seq{\Delta}{\cond{b}}}{\Par(\nops, A)}}{\pair{P \wedge \neg b}{P}}
	\]
	\justifies
	\hTriple{P}{\Fb(\Seq{\Seq{\Delta}{\cond{b}}}{\Par(\nops, A)})}{P \wedge \neg b}
	\using{(\Fb)}
\]
\justifies
{\bf \hTriple{P}{\while_b(A)}{P \wedge \neg b}}
\using{(\textup{def})}
\end{prooftree}
\]
\caption{Derivation of Hoare's while loop rule in $\Hsystem{\fLang}{\fcatS}{\fcH}$}
\label{while-derivation}
\end{figure*}

\subsection{Separation logic}
\label{separation}

In the case of the extended flowchart language $\pLang$ (pointer programs) and its respective semantics, we can define a verification functor $\ppH$ as follows. On the objects $X \in \pcatS$ we let $\ppH(X)$ be a pre-ordered set of formulas with enough primitives to describe the weakest pre-condition of programs, but without $\abort$ as an atomic formula (we want $\abort$ not to be expressible in the language). The ordering $P \iord R$ on the elements of $\ppH(X)$ is again taken to be $P \to R$ in the fixed theory. For each pointer program $A$ we let $\ppH(\sm{A})$ be the following monotone relation
\eqleft{\rel{P}{\ppH(\sm{A})}{Q} \,\fdefin\, P \to \wPre(A, Q)}
where $\wPre(A, P)$ is a formula expressing the weakest pre-condition of program $A$ under post-condition $Q$. As in Reynolds \citeyear{Reynolds:2002}, well-specified programs do not abort, since formulas in $\ppH(X)$ never hold true for $\abort$.

It has been shown in \cite{OHearn:2001,Reynolds:2002}, that the weakest liberal pre-conditions for the new basic statements can be concisely expressed in \emph{separation logic} as (see \cite{Reynolds:2002} for notation)
\eqleft{
\begin{array}{lcl}
	\wPre(x \assign [t], P) & \fdefin & \exists v' ((t \mapsto v') * ((t \mapsto v') \mwand P[v'/x])) \\[2mm]
	\wPre([t] \assign s, P) & \fdefin & (t \mapsto -) * ((t \mapsto s) \mwand P) \\[2mm]
	\wPre(x \assign {\sf new}(\vec{t}), P) & \fdefin & \forall i ((i \mapsto \vec{t}) \mwand P[i/x]) \\[2mm]
	\wPre({\sf disp}(t), P) & \fdefin & (t \mapsto -) * P
\end{array}}

Similarly to Lemma \ref{crucial} and Theorem \ref{while-trace}, one can show that the $\ppH$ defined above is a verification functor. The system $\Hsystem{\pLang}{\pcatS}{\ppH}$, which we then obtain from our abstract approach is basically the one presented in Reynolds \citeyear{Reynolds:2002}, for \emph{global backward reasoning}, using the logical theory of bunched implications as an `oracle' for the consequence rule.

\subsection{Hoare logic for running time analysis and termination}
\label{complexity}

Finally, we conclude this section with a new Hoare logic for \emph{running time} analysis (and \emph{termination}) of programs. For this we need a finer semantic model $\tfcatS$ for the flowchart programs. We need a model that distinguishes programs with different running time. Let us consider the identity operation and twist as neutral. We define the running time of a program to be the number of times non-neutral basic operations (e.g. assignment or conditional) are evaluated. Then two programs are identified if they have the same input-output behaviour and same running time on all terminating inputs. This gives rise to a refinement of the semantic model $\fcatS$ which we will call $\tfcatS$.

Let $\NN^\infty$ denote the usual pre-order of the natural numbers extended with $\infty$ as the top element, i.e. $n \leq \infty$ for all $n \in \NN$. Arithmetic with $\infty$ is done as usual, e.g. $n + \infty = \infty$. We want to use $\NN$ as a counter for the number of steps a program takes to terminate. Non-terminating executions are associated with $\infty$ which should be thought of as ``infinite time". For any set $\Sigma^m (\equiv \Sigma \uplus \ldots \uplus \Sigma)$ the pre-order on $\NN^\infty$ induces a pointwise pre-order on the set of functions $\Sigma^m \to \NN^\infty$. We take these pre-ordered sets $\{\Sigma^m \to \NN^\infty\}_{m \in \NN}$ as the objects of our category, with monotone relations between these as morphisms. This forms a subcategory of $\Hoa$ with cartesian product as the monoidal structure. 


Let $P, Q$ be functions in $\Sigma^n \to \NN^\infty$ and $\Sigma^m \to \NN^\infty$, respectively. Note that $P$ for instance can be viewed as an $n$-tuple $(P_1, \ldots, P_n)$ of functions $P_i : \Sigma \to \NN^\infty$. Given a flowchart $A$ (assume $\sm{A} : \Sigma^n \to \Sigma^m$) we define the following functor
\eqleft{\rel{P}{\tfcH(\sm{A})}{Q} \,\fdefin P \geq \RRT(A, Q)}
%
%
where $\RRT(A, Q) : \Sigma^n \to \NN^\infty$ calculates the \emph{relative running time} of $A$ with respect to $Q$. Intuitively, if $A$ does not terminate on input $\rho$ then $\RRT(A, Q)(\rho) = \infty$, otherwise $\RRT(A, Q)(\rho)$ calculates ${\sf RunTime}(A, \rho) + Q(\sm{A}(\rho))$. This can be given for the basic programs as
\eqleft{
\begin{array}{lcl}
	\RRT(\nops, Q) & \fdefin & Q \\[2mm]
	\RRT(x \assign t, Q) & \fdefin & Q[t/x]+1 \\[2mm]
	\RRT(\cond{b}, \pair{P}{Q}) & \fdefin & \ifthen{b}{P}{Q}+1 \\[2mm]
	\RRT(\twist, \pair{P}{Q}) & \fdefin & \pair{Q}{P} \\[2mm]
	\RRT(\Delta, Q) & \fdefin & \pair{Q+1}{Q+1}
\end{array}}
As usual, in order to obtain a completeness result we assume that the language over which we are defining our pre- and post-conditions is rich enough to express $\RRT$. Given that $\RRT(A, 0)(\rho) = \infty$ implies the program $A$ does not terminate on input $\rho$, we can conclude that any expressive language will need to include non-computable functions. This is to be expected, since an expressive language where each function is computable would in this case allow us to solve the halting problem. It is easy to see, however, that extending a Turing complete language with an oracle for the halting problem is sufficient for obtaining an expressive language.

\begin{figure*}[t]
\[ 
\begin{prooftree}
\[
	\[
		\[
			\[
				(\cond{b} \in \bLang_0)
				\justifies
				\hTriple{\ifthen{b}{P}{Q}+1}{\cond{b}}{\pair{P}{Q}}
			\]
			\[
				{\bf \hTriple{P}{A}{\ifthen{b}{P}{Q}+1}}
				\justifies
				\hTriple{\pair{P}{Q}}{\Par(\nops, A)}{\pair{P}{\ifthen{b}{P}{Q}+1}}
			\]
			\justifies
			\hTriple{\ifthen{b}{P}{Q}+1}{\Seq{\cond{b}}{\Par(\nops, A)}}{\pair{P}{\ifthen{b}{P}{Q}+1}}
		\]
		\justifies
		\hTriple{\pair{\ifthen{b}{P}{Q}+1}{\ifthen{b}{P}{Q}+1}}{\Seq{\Seq{\Delta}{\cond{b}}}{\Par(\nops, A)}}{\pair{P}{\ifthen{b}{P}{Q}+1}}
	\]
	\justifies
	\hTriple{\ifthen{b}{P}{Q}+1}{\Fb(\Seq{\Seq{\Delta}{\cond{b}}}{\Par(\nops, A)})}{P}
	\using{(\Fb)}
\]
\justifies
{\bf \hTriple{\ifthen{b}{P}{Q}+1}{\while_b(A)}{P}}
\using{(\textup{def})}
\end{prooftree}
\]
\caption{While loop rule for running time}
\label{while-run-derivation}
\end{figure*}

As in Definition \ref{abs-triple}, the functor $\tfcH$ gives rise to a Hoare triple $\hTriple{P}{A}{Q}$. The relation $\hTriple{P}{A}{Q}$ holds if and only if on each input $\rho$ such that $P(\rho) \neq \infty$ then $Q \neq \infty$ and $A$ on $\rho$ runs in time $P(\rho) - Q(\sm{A}(\rho))$ (this in particular implies that the program terminates).

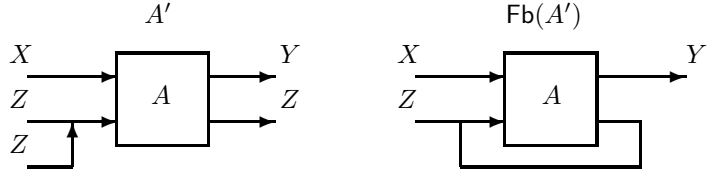
\begin{figure*}[t]
\begin{center}
\setlength{\unitlength}{6mm}
\begin{picture}(16,3.2)
	\thicklines
	\put(3.0,3){$A'$}
	\put(11.0,3){$\Fb(A')$}
	\put(0.4,1.8){\vector(1,0){2.0}}
	\put(0.4,0.8){\vector(1,0){2.0}}
	\put(0.4,-0.2){\line(1,0){1.0}}
	\put(1.4,-0.2){\vector(0,1){1.0}}
	\put(2.4,0.3){\framebox(2,2)}
	\put(3.15,1.2){$A$}
	\put(4.4,1.8){\vector(1,0){1.5}}
	\put(4.4,0.8){\vector(1,0){1.5}}
	\put(0.0,2.1){$X$}
	\put(0.0,1.1){$Z$}
	\put(0.0,0.1){$Z$}
	\put(6.0,2.1){$Y$}
	\put(6.0,1.1){$Z$}
	\put(9.0,1.8){\vector(1,0){2.0}}
	\put(9.0,0.8){\vector(1,0){2.0}}
	\put(11.0,0.3){\framebox(2,2)}
	\put(11.8,1.2){$A$}
	\put(13.0,1.8){\vector(1,0){2.0}}
	\put(13.0,0.8){\line(1,0){1.0}}
	\put(14.0,0.8){\line(0,-1){1.0}}
	\put(14.0,-0.2){\line(-1,0){4.0}}
	\put(10.0,-0.2){\line(0,1){1.0}}
	\put(8.6,2.1){$X$}
	\put(8.6,1.1){$Z$}
	\put(15.0,2.1){$Y$}
\end{picture}
\end{center}
\caption{Cook's construction (backward reasoning)}
\label{cook-trick-new}
\end{figure*}

\begin{theorem} $\tfcH : \tfcatS \to \Hoa$ as defined above is a verification functor.
\end{theorem}
{\bf Proof.} First of all, it is easy to check that $\tfcH$ indeed maps objects and morphisms of $\tfcatS$ into pre-orders and monotone relations (objects and morphisms of $\Hoa$). $\tfcH$ also trivially satisfies conditions (SC1) and (SC2), i.e. $\tfcH$ is a monoidal functor. One direction of condition (SC3) is similar to the proof of Lemma \ref{crucial}. Given a flowchart $A$ we construct a new flowchart $A'$ (see Figure \ref{cook-trick-new}) in such way that $\RRT(\Fb(A'), R)$ gives us the (greatest) fixed point of $A$. The other direction of condition (SC3) is similar to the proof of Theorem \ref{while-trace}.  \eProof

Therefore, this gives rise to a Hoare logic system $\Hsystem{\fLang}{\tfcatS}{\tfcH}$ which can be used to prove upper bounds on the running time of a program. As a simple example, consider the program calculating the factorial of $x$ described in Section \ref{system} (Figure \ref{factorial}):
\[ 
{\small 
\begin{prooftree}
\hTriple{3y+2}{x:=1}{3y+1}
\quad
\[
	\[
		\hTriple{3y}{x:=xy}{3y-1} \quad
		\hTriple{3y-1}{y:=y-1}{3y+1}
		\justifies
		\hTriple{3y}{x:=xy;y:=y-1}{3y+1}
	\]
	\justifies
	\hTriple{3y+1}{\while_b(x:=xy;y:=y-1)}{3y}
\]
\justifies
\hTriple{3y+2}{(x:=1);\while_b(x:=xy; y:=y-1)}{3y}
\end{prooftree}}
\]
where $b \equiv (y \geq 1)$. Given that $y = 0$ at the end, we have a proof of
\eqleft{\hTriple{3x-1}{(y:=x);\while_b(y:=y-1;x:=xy)}{0},}
which gives an upper bound of $3x - 1$ steps on the running time of the program.

\section{Verification Functors for Networks}
\label{network}

We will now apply our approach to Hoare logic to systems described by network diagrams that give a visual representation for sets of equations.
An important motivating example is the notion of signal flow graph in control theory which forms the basis of systems such as Simulink.
Network diagrams of this sort can be given several different formal semantics, e.g., in terms of differential equations or in terms of recurrence relations.

To apply our approach to network diagrams, our starting point is again the category of sets and relations, but now using the monoidal structure given by the cartesian product.
The {\em cartesian trace}, $\Tr(r)$, of a relation $r : \Arr{X \times Z}{Y \times Z}$ is defined by:
\eqleft{\rel{x}{\Tr(r)}{y} \;\pdefin\; \exists z (\rel{\pair{x}{z}}{r}{\pair{y}{z}}).}
In a network diagram, $r$ above will be thought of as the input-output relation of a subsystem, with $z$ representing a control signal that is fed back from output to input to form the overall system.
The input-output relation of the overall system is then the relation between $X$ and $Y$ determined by solving the system of equations denoted by $r$ for the value of the control signal, $z$.

Algebraic structure will be important to us. Fortunately, the cartesian trace gives rise to a wide range of interesting TMCs with nice algebraic properties.  In the sequel, we will first show that the cartesian trace applies to many interesting categories of relations enjoying useful algebraic or other structure.
We then present a diagrammatic language for stream circuits and give it a semantics in terms of an algebraic TMC $\scatS$ of relations representing a formal abstraction of the differential equation semantics.
We use a syntactic criterion to identify a useful class of ``valid'' circuits, whose semantic values are total functions.
The subcategory formed by these circuits admits a verification functor and so this gives a a sound and complete Hoare logic for the valid circuits.

\subsection{Traced Monoidal Categories of Relations}
\label{TMCsOfRelations}

Let $C$ be any concrete category, so that each object $X$ of $C$ has an associated {\em underlying set} (also written $X$ by abuse of notation) and each $C$-morphism between $C$-objects $X$ and $Y$ is a function from $X$ to $Y$, morphisms being composed via functional composition.
Let us assume that {\em(i)} set-theoretic finite products and equalizers\footnote{Recall that in any category, an equalizer for two morphisms $f, g : \Arr{X}{Y}$ is a universal arrow $h: \Arr{E}{X}$ making the horizontal composites in the diagram $E \stackrel{h}{\rightarrow} X \LowerStackRel{f, g}{\DArrow} Y$ equal. In the category of sets, an equalizer of any two maps $f, g : \Arr{X}{Y}$ is given by the inclusion of the subset $\{x : X \| f(x) = g(x)\}$, which we will call {\em the} equalizer of $f$ and $g$.} also serve as finite products and equalizers in $C$,
and {\em(ii)}, if $f$ is a $C$-morphism then the graph of $f$, as a subset of $X \times Y$, is the range of some $C$-morphism,
i.e., there is a $C$-object $U$ and a $C$-morphism $u:\Arr{U}{X \times Y}$, with $\Graph{f} = \Ran{u}$.
These assumptions hold, for example, for any concrete category defined by a set of operations and equational laws, such as any of the usual algebraic categories: groups, vector spaces over a field, modules over a ring etc.
In fact, for these categories, the converse of assumption {\em(ii)} holds: a function is a $C$-morphism iff its graph is the range of a $C$-morphism.
Under these assumptions, define a $C$-{\em relation} between $C$-objects $X$ and $Y$ to be any relation, $r$, which is given,as a subset of $X \times Y$, by the range of a $C$-morphism, i.e., $r = \Ran{u}$, for some $C$-object $U$ and $C$-morphism $u: \Arr{U}{X \times Y}$.
We then define the category $\CRel{C}$ to have the same objects as $C$ and to have as morphisms between $X$ and $Y$ the set of all $C$-relations between $X$ and $Y$, morphisms being composed by relational composition.
\begin{theorem}\label{algebraic-tmcs}
Under the above assumptions on the concrete category $C$, $\CRel{C}$ together with cartesian product and cartesian trace forms a TMC having $C$ as a subcategory.
If, moreover, a function is a $C$-morphism iff its graph is the range of some $C$-morphism, then this subcategory comprises precisely those relations in $\CRel{C}$ which are set-theoretic functions.
\end{theorem}
{\bf Proof.}
We must first verify that $\CRel{C}$ actually is a category:
by assumption {\em(i)} if $X$ is any $C$-object, then the diagonal function $\delta = x \mapsto \pair{x}{x}$ is a $C$-morphism from $X$ to $X \times X$
and $\Ran{\delta}$ is the identity relation on $X$ which is therefore a $\CRel{C}$-morphism acting as a two-sided identity for relational composition.
Given $\CRel{C}$-morphisms $r:\Arr{X}{Y}$ and $s:\Arr{Y}{Z}$, there are $C$-objects $U$ and $V$ and $C$-morphisms $u: \Arr{U}{X \times Y}$ and $v: \Arr{V}{Y \times Z}$ with $r = \Ran{u}$ and $s = \Ran{v}$.
Let us write $\pi_i: X_1 \times X_2 \to X_i$ for the projections of a binary product onto its factors and write $\pi_{ij}$ for the composite $(\pi_i; \pi_j)$.
Our data provide everything except the object $E$ and the dotted morphisms in the following diagram.
\[
\begin{diagram}
E & \rTo[dotted]^{i} & U \times V & \rTo^{u \times v}
	& (X \times Y) \times (Y \times Z)
		& \pile{\rTo^{\pi_{12}} \\ \rTo_{\pi_{21}}} & Y \\
   & \rdTo[dotted]^{e}(4,2) & &
	& \dTo_{\pi_1 \times \pi_2} & & \\
   & & &
	& {X \times Z}
\end{diagram}
\]

If we take $i:\Arr{E}{U}$ to be the equalizer of the two horizontal composites
from $U \times V$ to $Y$ and let $e:\Arr{E}{X \times Z}$ be the composite
$(i; u \times v;\pi_1 \times \pi_2)$
then 
$e$ is a $C$-morphism and $\Ran{e} = (r; s)$.
Thus $\CRel{C}$ has identities and is closed under relational composition and hence forms a category. It is easy to verify that the cartesian product makes $\CRel{C}$ into a symmetric monoidal category and, using assumption {\em(ii)}, that $C$ forms a subcategory, and that this subcategory coincides with the functional relations in $\CRel{C}$ under the stated condition.
So it remains to show $\CRel{C}$ is closed under cartesian trace.
So assume $r : \Arr{X \times Z}{Y \times Z}$ is a $\CRel{C}$-morphism, so that there is a $C$-object $U$ and a $C$-morphism $u: \Arr{U}{(X \times Z) \times (Y \times Z)}$ with $r = \Ran{u}$.
This gives everything except the object $E$ and the dotted morphisms in the following diagram.
\[
\begin{diagram}
E & \rTo[dotted]^{i} & U & \rTo^{u}
	& (X \times Z) \times (Y \times Z)
		& \pile{\rTo^{\pi_{12}} \\ \rTo_{\pi_{22}}} & Z \\
   & \rdTo[dotted]^{e}(4,2) & &
	& \dTo_{\pi_1 \times \pi_1} & & \\
   & & &
	& {X \times Y}
\end{diagram}
\]
If we take $i: \Arr{E}{U}$ to be the equalizer of the two horizontal composites from $U$ to $Z$ 
and let $e: \Arr{E}{X \times Y}$ be the composite $(i; u; \pi_1 \times \pi_1)$ then $\Tr(r) = \Ran{e}$ so the cartesian trace is a $\CRel{C}$-morphism and the proof is complete.
\eProof

As an aside, we note that this theorem offers an alternative to the partial  traced monoidal categories that have been considered by several researchers, e.g., see \cite{Haghverdi:2005}.  Rather than a partial trace on $C$, it gives a total trace on $\CRel{C}$ which contains $C$ as a subcategory.  Specific examples of partial TMCs given by Haghverdi and Scott \citeyear{Haghverdi:2005} are the category of vector spaces and the category of complete metric spaces with non-expansive continuous functions as morphisms. Both these examples are covered by the above theorem and without the restriction to finite dimensions in the case of vector spaces. The theorem also applies to both Hilbert spaces and Banach spaces with bounded operators as morphisms and to compact Hausdorff topological spaces with continuous maps.


Perhaps surprisingly, the dual of the construction in the above theorem also goes through using coequalizers, coproducts and an appropriate notion of cograph. Constructions of this sort have been considered in connection with universal algebra, see \cite{Hutchinson:1992} for details and references.

\subsection{The programming language: stream circuits}
\label{streams}

In \cite{Boulton:2003}, a Hoare logic was presented for the frequency analysis of linear control systems with feedback,  modelled as linear differential equations, an approach which we generalised and systematised in \cite{Arthan:2007}. In the present paper, we will model linear differential equations as stream circuits (see \cite{Rutten:2004}). With Theorem~\ref{algebraic-tmcs} in mind, this model fits nicely both with the algebraic viewpoint of \cite{Arthan:2007} and with the present framework.

\begin{figure*}[t]
\setlength{\unitlength}{8mm}
\begin{picture}(14,4)
	\thicklines
	\put(1.25,3.5){Identity}
	\put(1.25,3.0){\vector(1,0){1.5}}
	\put(6.5,4.0){Scalar}
	\put(5.5,3.0){\vector(1,0){1.0}}
	\put(6.5,2.5){\framebox(1.0,1)}
	\put(6.7,2.9){$a \times$}
	\put(7.5,3.0){\vector(1,0){1.0}}
	\put(10.5,3.5){Split}
	\put(10.0,3.0){\line(1,0){1.0}}
	\put(11.0,3.0){\vector(2,-1){1.0}}
	\put(11.0,3.0){\vector(2,1){1.0}}
	\put(1.3,1.5){Integrator}
	\put(0.5,0.5){\vector(1,0){1.0}}
	\put(1.5,0.0){\framebox(1.0,1)}
	\put(1.8,0.4){$\int$}
	\put(2.5,0.5){\vector(1,0){1.0}}
	\put(6.5,1.5){Twist}
	\put(6.0,1.0){\vector(2,-1){2.0}}
	\put(6.,0.0){\vector(2,1){2.0}}
	\put(10.7,1.5){Sum}
	\put(10.0,0.0){\vector(4,1){1.0}}
	\put(10.0,1.0){\vector(4,-1){1.0}}	
	\put(11.5,0.5){\circle{1.0}}
	\put(11.35,0.4){$+$}
	\put(12.0,0.5){\vector(1,0){1.0}}
\end{picture}
\vspace{2mm}
\caption{Basic stream circuits.}
\label{basicStream}
\end{figure*}
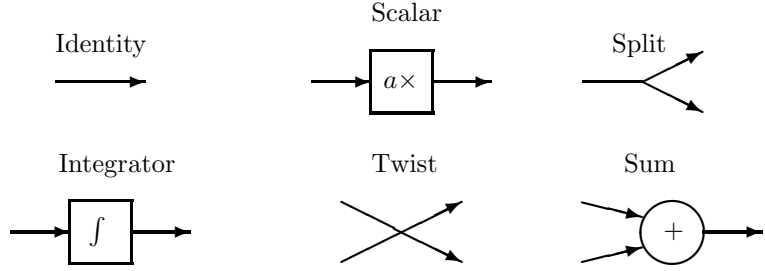

Our language of stream circuits comprises finite diagrams,  made up from the basic circuits shown in Figure \ref{basicStream} (where $a$ is a real number-valued parameter).
New circuits are formed from old by connecting outputs (arrow heads) to inputs (arrow tails).
An output of a circuit may be connected to an input of the same circuit, so we can create feedback loops.
Following \cite{Rutten:2004}, we say circuit is {\em valid} if every closed path in the circuit passes through at least one integrator (or ``register'' in Rutten's terminology).

\subsection{The semantics}

In our semantics, stream circuits denote differential equations.  As demonstrated in \cite{Escardo:1998}, many computational aspects of mathematical analysis can be developed in a co-algebraic setting using infinite streams of real numbers, an analytic function $f$ being modelled by the infinite stream $[f(0), f'(0), f''(0), \ldots]$ of its partial derivatives. For many purposes we may abstract away issues of convergence, and view infinite streams rather than functions as the main objects of interest.  Following \cite{Rutten:2004}, we can then view a stream circuit as just a system of equations on streams: for example, the differential equation $f = f'$, or equivalently $f = (\int f) + c$, corresponds to a stream equation $f = c \mathop{{::}} \sTl(f)$.

To turn these ideas into a formal semantics for stream circuits, let $\Sigma$ denote the set of streams of real numbers, i.e. all functions $\NN \to \RR$. Let the objects of $\scatS$ be the set containing the singleton set $\{\varepsilon\}$ and $\Sigma$, and closed under cartesian product, i.e. the objects are $\Sigma^0 = \{\varepsilon\}, \Sigma^1 = \Sigma, \Sigma^2 = \Sigma \times \Sigma, \ldots, \Sigma^m, \ldots$. Consider the following basic morphisms between the objects of $\scatS$: \\[2mm]
%
\begin{minipage}[t]{120mm}
\begin{minipage}[t]{55mm}
	\begin{itemize}
		\item {\it Wire}, $\nops : \Sigma \to \Sigma$
		\item[] \quad $\nops(\sigma) \fdefin \sigma$ \\[-2mm]
		\item {\it Scalars}, $(a \times) : \Sigma \to \Sigma$
		\item[] \quad $(a \times)(\sigma) \fdefin [a \sigma_0, a \sigma_1, \ldots]$ \\[-2mm]
		\item {\it Integrator}, $\sR : \Sigma \to \Sigma$
		\item[] \quad $\sR(\sigma) \fdefin [0, \sigma_0, \sigma_1, \ldots]$
	\end{itemize}
\end{minipage}
\begin{minipage}[t]{60mm}
	\begin{itemize}
		\item {\it Copy}, $\sCopy : \Sigma \to \Sigma \times \Sigma$
		\item[] \quad $\sCopy(\sigma) \fdefin \pair{\sigma}{\sigma}$ \\[-2mm]
		\item {\it Sum}, $(+) : \Sigma \times \Sigma \to \Sigma$
		\item[] \quad $(+)\pair{\sigma}{\sigma'} \fdefin [\sigma_0 + \sigma_0', \sigma_1 + \sigma_1', \ldots]$ \\[-2mm]
		\item {\it Twist}, ${\sf t} : \Sigma \times \Sigma \to \Sigma \times \Sigma$
		\item[] \quad ${\sf t}\pair{\sigma}{\sigma'} \fdefin \pair{\sigma'}{\sigma}$
	\end{itemize}
\end{minipage}
\end{minipage} \\[2mm]
The morphisms of $\scatS$ are the \emph{relations} obtained from the basic morphisms above viewed as relations (via their graph) and closing this set under relational composition, cartesian product of relations, and the trace for the cartesian product.

\begin{figure*}[h]
\begin{center}
	\setlength{\unitlength}{6.5mm}
	\begin{picture}(13,2.5)
		\thicklines
		\put(1.3,2.2){$\Tr(\sCopy \circ (+))$}
		\put(0.0,1.0){\vector(1,0){0.7}}
		\put(1.0,1.0){\circle{0.6}}
		\put(0.8,0.88){$+$}
		\put(1.3,1.0){\vector(1,0){3.5}}
		\put(4.0,1.0){\line(0,-1){1.5}}
		\put(4.0,-0.5){\line(-1,0){3.0}}
		\put(1.0,-0.5){\vector(0,1){1.2}}
		\put(7.5,2.4){$\Tr(((+) \times \id) \circ (\id \times \sCopy))$}
		\put(8.0,1.5){\vector(1,0){3.7}}
		\put(12.0,1.5){\circle{0.6}}
		\put(11.8,1.38){$+$}
		\put(12.3,1.5){\vector(1,0){0.5}}
		\put(12.0,0.5){\line(0,-1){1.0}}
		\put(12.0,-0.5){\line(-1,0){3.0}}
		\put(9.0,-0.5){\vector(0,1){1.0}}
		\put(9.0,0.5){\vector(1,0){3.0}}
		\put(12.0,0.5){\vector(0,1){0.7}}
	\end{picture}
\end{center}
\caption{Stream circuits not defining functions}
\label{pathological}
\end{figure*}
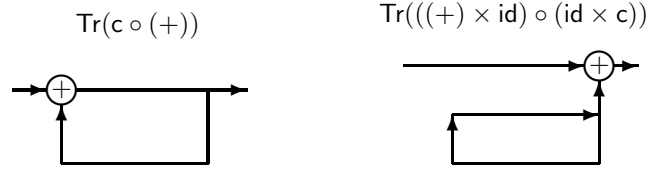

The diagrams in Figure \ref{pathological} illustrate two stream circuits that define relations that are not total functions. In the circuit on the left there is no fixed point for the input stream $\sigma_a \pdefin [a, a, \ldots]$ unless $a = 0$, so the circuit defines a partial function with domain $\{0\}$. On the other hand, in the circuit on the right any stream $\tau$ is a fixed point of the trace, so that any input stream $\sigma$ is related to any output stream, so the circuit represents the ``chaotic'' relation $\Sigma \times \Sigma$.
The notion of valid circuit gives a syntactic criterion for avoiding this pathological behaviour: while $\scatS$ is a category of relations that are partial and one-to-many in general, valid circuits denote \emph{total} functions, i.e.  a valid circuit denotes a system of equations whose solutions always exist and are unique (as we shall prove shortly). 


\begin{figure*}[t]
\begin{center}
\setlength{\unitlength}{6.5mm}
\begin{picture}(15,5)
	\thicklines
	\put(-0.8,4.3){${\sf fdback}(f) \pdefin \Tr((\sCopy) \circ (+) \circ (\nops \uplus f))$}
	\put(1.1,3.2){$\sCopy$}
	\put(2.8,2.68){$+$}
	\put(2.7,2.8){\vector(-1,0){1.5}}
	\put(1.2,2.8){\vector(-1,0){0.7}}
	\put(3.0,2.8){\circle{0.6}}
	\put(5.0,2.8){\vector(-1,0){1.7}}
	\put(3.5,0.8){\framebox(1,1)}
	\put(3.85,1.15){$f$}
	\put(1.2,2.8){\vector(0,-1){1.5}}
	\put(3.0,2.5){\line(0,-1){1.2}}
	\put(0.5,1.3){\line(1,0){0.7}}
	\put(3.5,1.3){\vector(-1,0){0.5}}
	\put(5.0,1.3){\vector(-1,0){0.5}}
	\put(0.5,0.3){\line(1,0){4.5}}
	\put(0.5,0.3){\line(0,1){1.0}}
	\put(5.0,0.3){\line(0,1){1.0}}
	\put(8.2,4.3){${\sf sum}(f, g) \pdefin (+) \circ (g \uplus f) \circ (\sCopy)$}
	\put(9.8,1.9){\vector(-1,0){0.8}}
	\put(10.1,2.2){\line(0,1){0.5}}
	\put(9.9,1.77){$+$}
	\put(10.1,1.9){\circle{0.6}}
	\put(10.1,1.6){\line(0,-1){0.5}}
	\put(10.8,2.7){\vector(-1,0){0.7}}
	\put(10.8,2.2){\framebox(1,1)}
	\put(11.15,2.6){$g$}
	\put(11.8,2.7){\line(1,0){0.7}}
	\put(12.5,1.9){\vector(0,1){0.8}}
	\put(10.8,1.1){\vector(-1,0){0.7}}
	\put(10.8,0.6){\framebox(1,1)}
	\put(11.15,0.95){$f$}
	\put(11.8,1.1){\line(1,0){0.7}}
	\put(12.5,1.9){\vector(0,-1){0.8}}
	\put(12.8,2.2){$\sCopy$}
	\put(13.3,1.9){\vector(-1,0){0.8}}
\end{picture}
\end{center}
\caption{Feedback and summation in $\scatS$}
\label{feedback}
\end{figure*}

It is again easy to see that $\scatS$ forms a symmetric monoidal category, with the monoidal structure of cartesian product. Finally, with the standard family of trace relations (defined in Section \ref{system}), $\scatS$ forms a traced symmetric monoidal category.
$\scatS$ has several important subcategories obtained by restricting the sets of morphisms: $\scatS^F$ in which the morphisms are the total functional relations, and $\scatS^V$ comprising the denotations of valid stream circuits.
To understand these subcategories it is helpful to consider some additional algebraic structure.
Let $\FPS = \RR[[x]]$ be the ring of formal power series with real coefficients, e.g., see~\citeN{Maclane:1999}, section VIII.9.
Thus elements of $\FPS$ are formal expressions $f_0 + f_1x + f_2x^2 + \ldots + f_mx^m + \ldots$ where the coefficients $f_i$ are real numbers.
Note that a univariate polynomial,  $p_0 + p_1x + p_2x^2 + \ldots + p_mx^m$, over $\RR$ may be viewed as a formal power series in which all but a finite number of the coefficients are zero.
Formal power series are added by adding corresponding coefficients and multiplied using the convolution product (generalising multiplication of polynomials), i.e., if $f = f_0 + f_1x + \ldots + f_mx^m + \ldots$ and $g = g_0 + g_1x + \ldots + g_mx^m \ldots$, the coefficients of the sum $f + g$ and the product $fg$ are given by:
\begin{eqnarray*}
(f + g)_m &=& f_m + g_m\\
(fg)_m &=& f_0g_m + f_1g_{m-1} + \ldots f_{m-1}g_1 + f_mg_0
\end{eqnarray*}
Under these operations $\FPS$ forms a commutative ring with $0 = 0 + 0x + \ldots$ and $1 = 1 + 0x + \ldots$.
If we identify a stream $\sigma$ with the formal power series $\sigma_0 + \sigma_1x^1 + \ldots + \sigma_mx^m + \ldots$, then $\FPS$ acts on the objects of $\scatS$ by multiplication, i.e., for $(\LSp{1}\sigma, \ldots, \LSp{n}\sigma) \in \Sigma^n$ and $f \in \FPS$, we define:
\begin{eqnarray*}
(\LSp{1}\sigma, \ldots, \LSp{n}\sigma)f = (\LSp{1}\sigma f, \ldots, \LSp{n}\sigma f)
\end{eqnarray*}
Under this action, the objects of $\scatS$ become $\FPS$-modules\footnote{%
Recall that the notion of $R$-module is the generalisation to arbitrary rings of the notion of vector space over a field: an $R$-module is an additive group $M$ such that $R$ acts on $M$ as a ring of linear operators. I.e., writing $x \mapsto xf$ for the action of $f \in R$ on $x \in M$, one has $(x + y)f = xf + yf$, $x1= x$ and $x(fg) = (xf)g$.
$R$ acts itself by right multiplication and so becomes an $R$-module whose submodules are called {\em ideals}, i.e., an ideal $I$ is an additive subgroup of $R$ such that, for any $a \in R$, $Ia = \{xa \| x \in I\} \subseteq I$.
See, e.g.,~\cite{Maclane:1999} chapter V for the basics of module theory.}.

If $f \in \FPS$ and $f_0 \not= 0$, then $f$ has a multiplicative inverse $f^{-1}$ whose coefficients are given recursively by
\begin{eqnarray*}
(f^{-1})_0 &=& 1/f_0\\
(f^{-1})_{m+1} &=& (f_1(f^{-1})_m + f_2(f^{-1})_{m-1} + \ldots + f_{m+1}(f^{-1})_0) / f_0
\end{eqnarray*}
as is readily checked.
Clearly, any non-zero $f \in \FPS$ may be written uniquely in the form $f = x^mg$ where $g$ is invertible, $m$ being the {\em order} of $f$, i.e., the smallest $m$ such that $f_m \not= 0$.
It follows that the non-zero ideals of $\FPS$ are precisely the principal ideals generated by the powers of $x$, i.e., $I \not = \{0\}$ is an ideal of $\FPS$ iff $I = x^m \FPS$ for some $m$.

It is easy to check that the basic morphisms of $\scatS$ are $\FPS$-module homomorphisms, so for example, if $\sigma \in \Sigma^n$ and $f \in \FPS$, one has $\sCopy(\sigma f) = \sCopy(\sigma)f$.
Noting that the approach of \cite{Arthan:2007} generalises to $\FPS$-modules\footnote{
The only difficulty is with theorem 5.8 of~\cite{Arthan:2007} where the proof in the published paper uses properties of bases to define relational inverse in terms of the feedback loop operator.
However, a basis-free construction can be given: in the notation of~\cite{Arthan:2007}, if $r:\Rel{V}{W}$, $r^{-1}$ is the composite:
\begin{eqnarray*}
(0, 1_W) &:& \Arr{W}{V \times W};\\
\Loop{1_{V \times W}}{1_{V \times W} - (\pi_1; r; (0, 1_W))} &:& \Rel{V \times W}{V \times W};\\
\pi_1 &:& \Arr{V \times W}{V}
\end{eqnarray*}
}, we find that each morphism of $\scatS$ is actually an additive relation, i.e., each morphism between objects $L$ and $M$ of $\scatS$ is a relation whose graph is a submodule of the product module $L \times M$.
Within $\scatS$, $\scatS^T$ comprises the subcategory of morphisms which happen to be total functions.
In the terminology of Section~\ref{TMCsOfRelations}, the methods of \cite[section 5]{Arthan:2007} show that $\scatS$ is the category of $\scatS^T$-relations: $\scatS = \CRel{\scatS^T}$

Let us exploit this algebraic structure to investigate the cartesian trace applied to a functional morphism in $\scatS$.
Using the vanishing law for the trace, the trace of a general morphism $\Sigma^m \times \Sigma^k \to \Sigma^n \times \Sigma^k$ can be calculated by iterating the trace on morphisms $\Sigma^{m'} \times \Sigma \to \Sigma^{n'} \times \Sigma$, so it suffices to consider the case $k = 1$.
So let $F: \Sigma^m \times \Sigma \to \Sigma^n \times \Sigma$ be a functional morphism, i.e., a homomorphism of $\FPS$-modules.
Just as with vector spaces, we can represent $F$ as a $2\times2$ block matrix, i.e., 
there are unique  $\FPS$-module homomorphisms, $A : \Sigma^m \to \Sigma^n$, $B : \Sigma \to \Sigma^n$, $C : \Sigma^m \to \Sigma$ and $D : \Sigma \to \Sigma$, such that $\pair{\alpha}{\sigma}F = \pair{\beta}{\tau}$ where:
\begin{eqnarray*}
\beta &=& \alpha A + \sigma B \\
\tau &=& \alpha C + \sigma D
\end{eqnarray*}
$D$ here is a $1 \times 1$ matrix, i.e., under the identification of $\Sigma$ with $\FPS$, $D$ is given by multiplication by some element of $\FPS$ say $D = (d)$.
The following lemma says that under appropriate assumptions on $d$, the trace of $F$ is a total function.
The ideal $I$ in the theorem may be thought of as a measure of the precision of an
approximation $\sigma$ to the actual fixed point $\sigma'$. For the simple existence of the fixed point, take $I = \FPS$ and pick an arbitrary $\sigma$.

\begin{lemma}\label{lemma:stream-circuit}
Assume $F : \Sigma^m \times \Sigma \to \Sigma^n \times \Sigma$ is defined by $\pair{\alpha}{\sigma}F = \pair{\beta}{\tau}$ where:
\begin{eqnarray*}
\beta &=& \alpha A + \sigma B \\
\tau &=& \alpha C + \sigma D
\end{eqnarray*}
for some  $A : \Sigma^m \to \Sigma^n$, $B : \Sigma \to \Sigma^n$, $C : \Sigma^m \to \Sigma$ and $D : \Sigma \to \Sigma$, where $D$ is given by the $1\times1$ matrix $(d)$ where $d \in \FPS$ is either 0 or has positive order, i.e., $d = xf$ for some $f \in \FPS$.
Let $I$ be any ideal in $\FPS$ and assume  $(\alpha, \sigma)F = (\beta, \tau)$ for some $\alpha$, $\beta$, $\sigma$ and $\tau$ where $\sigma - \tau \in I$, then there exist unique $\beta'$ and $\sigma'$ such that $(\alpha, \sigma')F = (\beta', \sigma')$ and then one has $\sigma' - \sigma \in I$.
\end{lemma}
{\bf Proof.}
For uniqueness, if $\beta'$ and $\sigma'$ satisfy
\begin{eqnarray*}
\beta' &=& \alpha A + \sigma' B \\
\sigma' &=& \alpha C + \sigma' D
\end{eqnarray*}
then, as $D = (d)$ and $1 - d$ is invertible, the second equation gives $\sigma' = \alpha C / (1 - d)$ fixing $\sigma'$ and then the first equation fixes $\beta'$. For the existence, if $I = \{0\}$, then $\tau = \sigma$ and we may take $\beta' = \beta$ and $\sigma' = \sigma$, otherwise define sequences $\LSp{m}\sigma$ and $\LSp{m}\tau$ of elements of $\FPS$ as follows:
\[\begin{array}{rclrcl}
\LSp{0}\sigma &=& \sigma, & \LSp{0}\tau &=& \tau = \alpha C + \LSp{0}\sigma d\\
\LSp{m+1}\sigma &=& \LSp{m}\tau, \quad&\quad \LSp{m+1}\tau &=& \alpha C + \LSp{m+1}\sigma d
\end{array}
\]
We claim that $\LSp{m}\sigma - \LSp{m}\tau \in Id^m$ for each $m$.
This is true by assumption for $m = 0$, so assume inductively that it holds for some $m$,
we then have
\begin{eqnarray*}
\LSp{m+1}\sigma - \LSp{m+1}\tau &=& \LSp{m}\tau - (\alpha C + \LSp{m}\tau d) \\
	&=& \LSp{m}\tau(1 - d) - \alpha C\\
	&=& \LSp{m}\tau(1 - d) - \LSp{m}\tau + \LSp{m}\sigma d\\
	&=& (\LSp{m}\sigma - \LSp{m}\tau) d \in (Id^m)d = Id^{m+1}
\end{eqnarray*}
which proves the claim.
Note that as $d = xf$ for some non-zero $f$, $\LSp{m}\sigma_k = \LSp{m+1}\sigma_k$ for $m > k$ since we have $\LSp{m+1}\sigma - \LSp{m}\sigma = \LSp{m}\tau - \LSp{m}\sigma \in Id^m = If^mx^m$. It follows that if we put $\sigma'_k = \LSp{k+1}\sigma_k$, then $\sigma' - \LSp{m}\sigma \in Id^m$ for all $m$ and if we put $\tau' = \alpha C + \sigma' D$ then $\sigma' - \tau' \in Id^m$ for all $m$, but the intersection of the ideals $Id^m$ is the zero ideal, so $\sigma' - \tau' = 0$, i.e., $\sigma' = \tau' = \alpha C + \sigma' D$; taking $\beta' = \alpha A + \sigma' B$ completes the proof.
\eProof
Remark: the solution in the case where $d = 0$, can also be given directly as $\sigma' = \alpha C$ and $\beta' = \alpha (A + CB)$.

\begin{theorem}\label{thm:valid-streams}
Valid stream circuits denote total functions.
\end{theorem}
By the lemma, it suffices to show that if a circuit denotes a function $F : \Sigma^m \times \Sigma \to \Sigma^n \times \Sigma$ given by $\pair{\alpha}{\sigma}F = \pair{\beta}{\tau}$ where:
\begin{eqnarray*}
\beta &=& \alpha A + \sigma B \\
\tau &=& \alpha C + \sigma D
\end{eqnarray*}
$D = (d)$ being a $1\times1$ matrix, and if every path in the circuit from the second input component to the second output component passes through an integrator, then $d = xf$ for some $f \in \FPS$.
This is easy to verify by induction on the construction of the circuit, since passing a stream through an integrator corresponds to multiplying it by $x$.
\eProof

\subsection{Hoare logic for stream circuits}

We will now construct a functor $\scH : \scatS \to \Hoa$ and show that its restriction to valid circuits is a verification functor. Each
object of $\scatS$ has the form $\Sigma^m = \Sigma \times \ldots \times
\Sigma$ for some $m$.  We define $\scH(\Sigma)$ to be the set of pairs $(\sigma,
I)$ where $\sigma \in \Sigma$ and $I \subseteq \Sigma$ corresponds to an ideal
in $\FPS$ under the identification of $\Sigma$ with $\FPS$.  $\scH(\Sigma)$ is
ordered by defining $\pair{\sigma}{I} \iord \pair{\tau}{J}$ iff $\sigma + I \subseteq
\tau + J$.  $\scH(\Sigma^m)$ is then defined to be $\scH(\Sigma)^m$ which we may
identify with  the set of all pairs $\pair{\alpha}{L}$ where $\alpha \in \Sigma^m$
and $L$ is what we may call a {\em coordinate} submodule of $\Sigma^m$, i.e.,
a submodule of the form $I_1 \times I_2 \times \ldots \times I_m$ where each
$I_k$ is an ideal of $\FPS$ under the identification of $\Sigma$ with $\FPS$.
Note that the $m$-tuple $\langle I_1, I_2, \ldots, I_m\rangle$ can be recovered
from the product $I_1 \times I_2 \times \ldots \times I_m$ since the ideals $I_j$ are non-empty sets.
The product ordering on $\scH(\Sigma)^m$ corresponds to taking $\pair{\alpha}{L} \iord \pair{\beta}{M}$ iff $\alpha + L \subseteq \beta + M$.
Thus we let $\pair{\alpha}{L}$ correspond to the subset $\alpha + L$ of $\Sigma^m$
and think of $\alpha$ as an estimate of a value in $\Sigma^m$ with $L$ giving the precision of the estimate.  So for example, $\pair{\alpha}{\Sigma^m}$ represents the whole set, or a maximally imprecise estimate, while $\pair{\alpha}{0}$ represents the singleton set $\{\alpha\}$, or an exact estimate.

On the morphisms (semantic values of stream circuits) $f : X \to Y$ in
$\scatS$, we define the relation $\scH(f)$ as follows, where $\pair{\alpha}{L} \in
\scH(X)$ and $\pair{\beta}{M} \in \scH(Y)$, and where $(A)r$ denotes the image of the
set $A$ under the relation $r$.
\eqleft{\rel{\pair{\alpha}{L}}{\scH(f)}{\pair{\beta}{M}} \fdefin \alpha + L \subseteq \Dom{f} \land (\alpha + L)f \subseteq \beta + M}
We now show that $\scH$ as defined above yields a verification functor for valid circuits.

\begin{theorem} \label{main-stream} $\scH : \scatS^V \to \Hoa$, as defined above, is a verification functor for valid circuits.
\end{theorem}
{\bf Proof.}
That $\scH$ commutes with composition of morphisms is easy to verify, so $\scH$ is indeed a functor.
By construction, $\scH(X \times Y) = \scH(X) \times \scH(Y)$
(although in our notation we are identifying
$(\Sigma \times \PP\Sigma)^m$ with a subset of $(\Sigma^m \times \PP(\Sigma^m$)),
and one may check that $\scH(f \times g) = \scH(f) \times \scH(g)$ when
$£$ and $g$ are morphisms, so $\scH$ is a monoidal functor. It remains to show that $\scH$ commutes with the trace.  As in Theorem~\ref{thm:valid-streams}, it suffices to consider
traces of morphisms $f: \Sigma^m\times\Sigma \to \Sigma^n \times \Sigma$.
So assume $\rel{\pair{\alpha}{L}}{\scH(\Tr(f))}{\pair{\beta}{M}}$, then by the definition
of $\scH$, $\alpha + L \subseteq \Dom{\Tr(f)}$ and so in particular, $\alpha \in
\Dom{\Tr(f)}$ which means that there is $\beta'\in \Sigma^n$ and $\sigma \in
\Sigma$, such that $\rel{\pair{\alpha}{\sigma}}{f}{\pair{\beta'}{\sigma}}$.  But
then from the definition of $\scH$ we must have that $\beta' - \beta \in M$ and
that
\eqleft{\rel{\pair{\pair{\alpha}{\sigma}}{L \times 0}}{\scH(f)}{\pair{\pair{\beta}{\sigma}}{M \times 0}}}
which means that $\scH(\Tr(f)) \subseteq \Tr(\scH(f))$.
Conversely, assume that
\eqleft{\rel{\pair{\alpha}{L}}{\Tr(\scH(f))}{\pair{\beta}{M}},}
so that there is a stream $\sigma$ and an ideal $I$ such that
\eqleft{\rel{(\pair{\pair{\alpha}{\sigma}}{L \times I}}{\scH(f)}{\pair{\pair{\beta}{\sigma}}{M \times I}}.}
Then $\pair{\alpha}{\sigma} \in \Dom{f}$, so that there are $\gamma$ and $\tau$
such that $\pair{\alpha}{\sigma}f = \pair{\gamma}{\tau}$, where $\gamma - \beta \in M$ and $\sigma - \tau \in I$.
Now, just as in Theorem~\ref{thm:valid-streams} the assumption that the
circuit is valid lets us apply Lemma~\ref{lemma:stream-circuit} to give
unique $\delta$ and $\sigma'$ such that $\pair{\alpha}{\sigma'}f = \pair{\delta}{\sigma'}$
such that $\delta - \gamma \in M$ and $\sigma' - \sigma \in I$,
but then also $\delta - \beta \in M$ and we may check that $\rel{\pair{\alpha}{L}}{\scH(\Tr(f))}{\pair{\beta}{M}}$.
Thus $\Tr(\scH(f)) \subseteq \scH(\Tr(f))$ completing the proof.
\eProof


This gives rise to a sound and complete Hoare-logic system for reasoning about \emph{valid} stream circuits. Notice that the rules are only sound for valid circuits.
For instance, consider the circuit $\sCopy \circ (+) :\Sigma \to \Sigma \times \Sigma$ used to construct the example shown on the left of Figure \ref{pathological}.
For any $\alpha$, the triple $\hTriple{\pair{\pair{\alpha}{0}}{\Sigma \times \Sigma}}{\sCopy \circ (+)}{\pair{0}{\Sigma \times \Sigma}}$ is valid,but the triple $\hTriple{\pair{\alpha}{\Sigma}}{\Tr(\sCopy \circ (+))}{\Sigma}$ is only valid for $\alpha = 0$

\begin{figure*}[h]
\[ 
\begin{prooftree}
\[
	\[
		\[
			\[
				((+) \in \fcatS)
				\justifies
				\hTriple{\pair{s}{t}}{(+)}{s+t}
			\]
			{\bf \hTriple{s+t}{f}{t}}
			\justifies
			\hTriple{\pair{s}{t}}{(+); f}{t}
			\using{({\sf Seq})}
		\]
		\[
			((\sCopy) \in \fcatS)
			\justifies
			\hTriple{t}{(\sCopy)}{\pair{t}{t}}
		\]
		\justifies
		\hTriple{\pair{s}{t}}{(+); f; (\sCopy)}{\pair{t}{t}}
		\using{({\sf Seq})}
	\]
	\justifies
	\hTriple{s}{\Tr((+); f; (\sCopy))}{t}
	\using{(\Fb)}
\]
\justifies
{\bf \hTriple{s}{{\sf fdback}(f)}{t}}
\using{(\textup{def})}
\end{prooftree}
\]
\caption{Derivation of feedback rule in $\Hsystem{\Lang}{\scatS}{H}$}
\label{feedback-derivation}
\end{figure*}

Streams can be viewed as giving the Taylor expansions of analytic functions. in this particular example, the pre- and post-conditions in the Hoare logic correspond to partial sums for the Taylor expansions. A partial sum comprising $n$ terms corresponds to the specification $\alpha + \Sigma x^{n}$, where $\alpha$ is any stream with $\alpha_i = s_i$ for $i = 0, \ldots, n - 1$.
We have thus obtained from our general categorical construction a sound and complete formal system $\Hsystem{\Lang}{\scatS}{\scH}$, for reasoning about valid stream circuits and their input-output behaviour over classes of functions with a common partial Taylor sum.

If the simple feedback circuit ${\sf fdback}(f)$ is defined in $\scatS$ by $\Tr((+); f; (\sCopy))$, we then have the following rule, whose derivation is given in Figure \ref{feedback-derivation} and is very similar to the derivation of the rule for while loops (cf. Figure~\ref{while-derivation}).
\[ 
\begin{prooftree}
{\bf \hTriple{s+t}{f}{t}}
\justifies
{\bf \hTriple{s}{{\sf fdback}(f)}{t}}
\end{prooftree}
\]

\subsection{Hilbert Spaces}

We have already remarked that the constructions of Section \ref{TMCsOfRelations} apply to the category of Hilbert spaces and bounded operators. This gives a TMC whose objects are Hilbert spaces and whose morphisms are relations whose graphs are complete subspaces of a product space.  Restricting to a specific Hilbert space $\Sigma$ and its finite products $\Sigma^m$, analogues of all our results on stream circuits go through using closed discs, $I_t = \{\sigma:\Sigma \| \Norm{\sigma} \leq t\}$ in place of ideals of $\FPS$.
The analogue of~\ref{lemma:stream-circuit} requires $D$ to be a contraction mapping ($\Norm{D} = \mbox{{\sf sup}}_{\sigma}(\Norm{\sigma D}/\Norm{D}) < 1$) and then the fixed point can be given as an analytic limit. For the lemma to serve its purpose, the hypothesis on the approximate fixed point needs to be slightly stronger than for stream circuits, but the assumptions available where the lemma is used in Theorem~\ref{main-stream} are sufficient.

The Hoare logic obtained in this way for systems where the semantic domains are Hilbert spaces way is interesting to compare with the Hoare logic for vector space semantics in general given in~\cite{Arthan:2007}. In the general approach we do indeed get a very general Hoare logic but at the price of rules involving side-conditions that have a non-trivial semantic content.
The approach for Hilbert spaces sketched here admits a much less general form of assertion and applies to a restricted category of spaces but those restrictions mean that it falls within the general framework of the present paper and yields rules that are much more satisfactory from a syntactic point of view.

\section{Conclusion and Related Work} 
\label{conclusion}

Several abstractions of the Floyd-Hoare logic can be found in the literature. In this final section we attempt to clarify the relation between the work presented here, and the previous work. Due to the vast amount of research in the area, we will only cover, however, works that we believe are closely related to ours.

Kozen's \citeyear{Kozen:2000} \emph{Kleene Algebra with Test}, KAT, consists essentially of a Kleene algebra with a Boolean subalgebra. 
In Kozen's work Hoare triples $\hTriple{P}{A}{Q}$ are modelled as equations $P A = P A Q$, using the multiplication available in KAT. The rules of Hoare logic are then obtained as consequences of the equational theory of KAT. Although our work is based on similar ideas (reducing Hoare triples to pre-order statements) there does not seem to be at the moment a clear cut connection between the two approaches. Whereas Kozen relies on the rich theory of KAT to derive the usual rules of Hoare logic, in our development we use a minimal theory of pre-ordered sets for obtaining soundness and completeness.

Kozen's work is related to previous work on iteration theory \cite{Bloom:1991,Arbib:1986} and dynamic logic \cite{Pratt:1976}. It should be stressed that all these focus on the semantics of Hoare logic over flowcharts and while programs, where the intrinsic monoidal structure is \emph{disjoint union}. As we have shown in Section \ref{network},  our approach is more general including systems with an underlying \emph{cartesian structure} as well.

Abramsky et al. \citeyear{Abramsky:1996} have also studied the categorical structure of Hoare logic, using the notion of \emph{specification structures}. It is easy to see that a TMC $\catS$ together with a verification functor $H : \catS \to \Hoa$ gives rise to a specification structure: $H$ maps objects $X \in \catSo$ to sets $H(X)$, and $H(f)(P) \iord Q$ defines a ternary relation $H(X) \times \catS(X, Y) \times H(Y)$. The extra structure of pre-order and trace, however, allows us to prove an abstract completeness theorem, which does not seem to be the focus of \cite{Abramsky:1996}.

Blass and Gurevich \citeyear{Gurevich:2000} considered the \emph{underlying logic of Hoare logic}. Since Cook's celebrated completeness result \cite{Cook:1978}, Cook-expressive first-order logics have been used in proofs of relative completeness for Hoare logic. Blass and Gurevich have shown that existential fixed-point logic $\EFL$ is sufficient for proving Cook's completeness result, without the need for Cook's expressiveness condition. $\EFL$ contains the necessary constructions to ensure that the functor $\fcH(f)$ of Section \ref{while} can be inductively built, rather than assumed to exist. The fixed-point construction is used in order to produce the fixed point $Q$ of Lemma \ref{crucial}.


For a given semantic mapping $\sm{\cdot} : \Lang \to \catS$, verification functor $H : \catS \to \Hoa$ and program $A : X \to Y$ in $\Lang$, the binary monotone relation $\hTriple{\cdot}{A}{\cdot}$, i.e.
\eqleft{\rel{(\cdot)}{H(\sm{A})}{(\cdot)} \quad : \quad H(X) \times H(Y)}
can be viewed as an order-preserving function $H(X)^{\rm op} \times H(Y) \to \BB$, where $H(X)^{\rm op}$ denotes the pre-order $H(X)$ with the opposite ordering, and the usual ordering for the product and the booleans $\BB$ is assumed. If we generalise pre-orders to categories, and the booleans $\BB$ to an arbitrary category, the binary relation $\hTriple{\cdot}{A}{\cdot}$ becomes a profunctor (cf. \cite{Winskel:2004}). Such profunctors also arise from a generalisation of relation, so-called {\em span}, used by Winskel \citeyear{Winskel:2005} to represent processes. A span is a diagram of the form $A \leftarrow U \rightarrow B$, where $A, B$ and $U$ are event structures (partial orders endowed with a binary ``consistency" relation). Such a diagram is equivalent to a morphism $U \to A \times B$ as used in our definition of the category of $C$-relations. These observations suggest several possible links between our methods and Winskel's approach to non-determinism and concurrency.



\begin{acks}
We are grateful for discussions we have had with Martin Hyland, and his suggestion of using monotone relations (rather than functions) which considerably simplified our previous work \cite{Martin:2006}. We also thank Glynn Winskel for helpful correspondence and pointers to related work. We thank Colin O'Halloran of QinetiQ and John Hall of DSTL for a discussion which stimulated our interest in applying Hoare logics to systems expressed as network diagrams.
\end{acks}

\bibliographystyle{acmtrans}


\begin{received}
\end{received}

\end{document}